\documentclass[12pt]{iopart}
\usepackage{graphicx}
\eqnobysec
\renewcommand{\theequation}{\thesection.\arabic{equation}}

\begin{document}

\title[Multi-soliton solutions of KP equation with integrable boundary]{Multi-soliton solutions of KP equation with integrable boundary via $\overline\partial$-dressing method}
\author{V. G. Dubrovsky, A. V. Topovsky}
\address{Novosibirsk State Technical University, Karl Marx prospect 20, 630072, Novosibirsk, Russia.}
\ead{\mailto{dubrovsky@ngs.ru},\mailto{dubrovskij@corp.nstu.ru}, \mailto{topovskij@corp.nstu.ru}, \mailto{topovsky.av@gmail.com}}

\begin{abstract}
New classes of exact multi-soliton solutions of KP-1 and KP-2 versions of Kadomtsev-Petviashvili equation with integrable boundary condition $u_{y}\big|_{y=0}=0$ by the use of $\overline\partial$-dressing method of Zakharov and Manakov are constructed in the paper. General determinant formula in convenient form for such solutions is derived. It is shown how reality and boundary conditions for the field $u(x,y,t)$ in the framework of $\overline\partial$-dressing method can be satisfied exactly. Explicit examples of two-soliton solutions as nonlinear superpositions of two more simpler \,"deformed"\, one-solitons are presented as illustrations: the fulfillment of boundary condition leads to formation of bound state of two more simpler one-solitons, resonating eigenmodes of $u(x,y,t)$ in semi-plane $y\geq0$ as analogs of standing waves on the string with fixed end points.
\end{abstract}

\noindent{\it Keywords\/}: Kadomtsev-Petviashvili equation,  $\overline{\partial}$-dressing method, exact multi-soliton solutions,  integrable boundary condition
\pacs{02.30.Ik, 02.30.Jr, 02.30.Zz, 05.45.Yv}

\section{Introduction}
\label{Section_1}
\setcounter{equation}{0}
\setcounter{figure}{0}
The famous Kadomtsev-Petviashvili  (KP) equation~\cite{KadPetv}-\cite{AblowitzClarkson}:
\begin{equation}\label{KP}
u_{t}+ u_{xxx}+6uu_x+3\sigma^2\partial_x^{-1}u_{yy}=0,
\end{equation}
with $\sigma=i$ for KP-1 and $\sigma=1$ for KP-2, can be represented as compatibility condition in the well known Lax form $\left[L_1,L_2\right]=0$ of two linear auxiliary problems~\cite{Dryuma},~\cite{ZakharovShabat}:
\begin{equation}\label{KP auxiltary problems}
\left\{
\begin{array}{ll}
L_1\psi=(\sigma\partial_y+\partial^2_x+u)\psi=0, \\
L_2\psi=(\partial_t+4\partial^3_x+6u\partial_x+3u_x-3\sigma^2\partial^{-1}_xu_y)\psi=0.
\end{array}
\right.
\end{equation}
The first linear problem (\ref{KP auxiltary problems}) with $y$ variable as \,"time"\, represents: nonstationary Schroedinger equation - for KP-1 case with $\sigma=i$, diffusion or heat equation - for KP-2 case with $\sigma=1$.

KP-equation (\ref{KP}) is the first discovered remarkable integrable example from now known long list of 2+1-dimensional integrable nonlinear equations.
 This equation has been well studied: several classes of exact solutions, hamiltonian and recursion structures, Cauchy problem, etc. have been analyzed by different methods see examples in ~\cite{Dryuma}-\cite{KonopelchenkoBook2}.

In the present paper new classes of exact multi-soliton solutions of KP equation (\ref{KP}) with integrable boundary condition \cite{Habibullin}, ~\cite{Habibullin2}
\begin{equation}\label{BoundaryCondition}
u_y(x,y,t)\big|_{y=0}=\frac{\partial u(x,y,t)}{\partial y}\bigg|_{y=0}=0
\end{equation}
for both versions KP-1 and KP-2 are constructed via $\overline\partial$-dressing method of Zakharov and Manakov~\cite{Manakov}-\cite{Beals&Coifman2}.
The concept of integrable boundary conditions compatible with integrability for integrable nonlinear equations was first introduced in the paper of Sklyanin~\cite{Sklyanin}. In subsequent works of Habibullin et al ~\cite{Habibullin}-\cite{HabibullinKDV} and others ~\cite{Vereshchagin} this concept to several of integrable nonlinear equations has been applied for different types of equations: difference equations, 1+1-dimensional and 2+1-dimensional integrable nonlinear differential and integro-differential  equations; a list of integrable boundary conditions for known 2+1-dimensional nonlinear equation such as KP, mKP, Nizhnik-Veselov-Novikov, Ishimori, Davey-Stewartson and so on with some examples of corresponding solutions have been proposed and calculated~\cite{Habibullin}-\cite{Vereshchagin}.
A. S. Fokas et al obtained interesting results via so called Unified Approach to Boundary Value problems, see book \cite{FokasBook}, where was demonstrated the applicability of this method for one-dimensional and multi-dimensional linear and nonlinear differential equations.

The construction of exact multi-soliton solutions of 2+1-dimensional integrable nonlinear equations with integrable boundaries also can be effectively done by powerful $\overline\partial$-dressing method of Zakharov and Manakov~\cite{Manakov}-\cite{Beals&Coifman2}.
This was demonstrated at first in the paper of Dubrovsky,  Topovsky and Ostreinov~\cite{Dubrovsky&Topovsky&OstreinovKP} where KP-2 equation with integrable boundary condition $(u_{xx}+\sigma u_{y})(x,y,t)\big|_{y=0}=0$  being considered.

In the present paper the calculations ~\cite{Dubrovsky&Topovsky&OstreinovKP} via $\overline\partial$-dressing of exact solutions for KP equation with another integrable boundary \cite{Habibullin}, ~\cite{Habibullin2}, i.e. $u_{y}\big|_{y=0}=0$, are continued.  It is shown how the boundary condition (\ref{BoundaryCondition}) and condition of reality $u=\overline u$ in construction of exact multi-soliton solutions of KP equation (\ref{KP}) can be effectively satisfied in the framework of $\overline\partial$-dressing method exactly by direct calculations with determinant forms of exact solutions. The restrictions from reality and integrable boundary conditions (obtained by the use of \,"limit of weak forms"\,)  are  also are applied but final results are checked in general explicit form with determinant formulas for exact solutions.

The paper is organized as follows. The first section is Introduction. In second section we reviewed the basic formulae of $\overline\partial$-dressing for KP equation (\ref{KP}), derived general determinant formula in convenient form for KP equation and the restrictions from reality and boundary conditions in the limit of weak fields.  In the following third and fourth sections new classes of multi-soliton solutions for KP-1 and KP-2 versions of KP equation are constructed and illustrated by the simple examples of two-soliton solutions. It is shown that imposition of integrable boundary condition $u_y\big|_{y=0}=0$ leads to formation of eigenmodes of field $u(x,y,t)$ in semi-plane $y\geq0$, or to certain bound state of several coherently connected with each other one-solitons. These eigenmodes of $u(x,y,t)$, propagating along $x$-axis with some velocity, resemble the standing waves on elastic string, arising from corresponding boundary conditions at endpoints of a string.

\section{Basic formulas of $\overline\partial$-dressing for KP equation, restrictions from reality and boundary conditions, determinant formula for exact multi-soliton solutions}
\label{Section_2}
\setcounter{equation}{0}
\setcounter{figure}{0}
First general formulas of $\overline\partial$-dressing method of Zakharov and Manakov in applications for KP equation are reviewed ~\cite{Manakov}-\cite{Bogdanov&Manakov}. Central object of $\overline\partial$-dressing is wave function $\chi(\lambda,\overline\lambda; x,y,t)$ which is the function of spectral variables $\lambda$, $\overline\lambda$ and spacetime variables $x,y,t$. This function is connected with wave function $\psi(\lambda,\overline\lambda; x,y,t)$ of linear auxiliary problems (\ref{KP auxiltary problems}) for KP (\ref{KP}) by the formula \cite{KonopelchenkoBook1}, \cite{KonopelchenkoBook2}:
\begin{equation}\label{Psi&Chi}
  \Psi(\lambda,\overline\lambda; x,y,t):=\chi(\lambda,\overline\lambda; x,y,t)\exp{F(\lambda;x,y,t)}.
\end{equation}
with phase $F(\lambda;x,y,t)$ in exponent
\begin{equation}\label{F(lambda)}
  F(\lambda; x,y,t)=i\lambda x+\frac{\lambda^2}{\sigma}y+4i\lambda^3t.
\end{equation}
$\sigma=i$ - for KP, $\sigma=1$ - for KP-2.
Herein  full notations $\chi(\lambda,\overline\lambda; x,y,t)$, $F(\lambda;x,y,t)$, $R_0(\mu,\overline\mu;\lambda,\overline\lambda)$, etc. as short notations are indicated with restriction to corresponding dependence on spectral variables $\lambda$, $\overline\lambda$ (or more shorter simply on $\lambda$) as $\chi(\lambda,\overline\lambda)$ or $\chi(\lambda)$, $F(\lambda)$; $\psi(\lambda,\overline\lambda)$ or $\psi(\lambda)$, $R_0(\mu,\lambda)$, respectively.

Basic equation of $\overline\partial$-dressing method is the so called $\overline\partial$-problem~\cite{KonopelchenkoBook1},~\cite{KonopelchenkoBook2} or equivalent to it singular integral equation for wave function $\chi(\lambda,\overline\lambda)$:
\begin{equation}\label{di_problem1}
\fl\chi (\lambda) = 1 + \frac{2i}{\pi}\int\int\limits_C {\frac{d{\lambda }'_R
d\lambda'_I}{(\lambda'-\lambda)}}
\int\int\limits_C  \chi(\mu,\overline{\mu})
R_0(\mu ,\overline \mu ;\lambda',\overline {\lambda' })e^{F(\mu)-F(\lambda')}{d\mu_R  d\mu _I}
\end{equation}
with kernel $R_0(\mu,\overline\mu;\lambda,\overline\lambda)$ (in short $R_0(\mu,\lambda)$). In (\ref{di_problem1}) due to (\ref{KP auxiltary problems}), (\ref{F(lambda)}), (\ref{Psi&Chi}) the canonical normalization of wave function $\chi\big|_{\lambda\rightarrow\infty}\rightarrow1$ is possible and will be further used in our paper, as usual $\lambda=\lambda_R+i\lambda_I$, $\mu=\mu_R+i\mu_I$ are the notations for complex spectral variables. Reconstruction formula for solution $u(x,y,t)$ of KP equation (\ref{KP})
\begin{equation}\label{reconstruct}
u(x,y,t)=-2i\partial_x \chi_{-1}(x,y,t)
\end{equation}
expresses $u$ through the $\chi_{-1}$ coefficient of Taylor expansion of $\chi(\lambda,\overline\lambda)$ in the neighborhood of $\lambda=\infty$:
\begin{equation}\label{TaylorExp}
\chi(\lambda,\overline\lambda)=1+\frac{\chi_{-1}}{\lambda}+\ldots \quad.
\end{equation}
Formula (\ref{reconstruct}) is valid for both versions of KP equation(\ref{KP}):  KP-1 and KP-2.  From (\ref{di_problem1}) one has for the coefficient $\chi_{-1}$ of (\ref{TaylorExp}):
\begin{equation}\label{chi_(-1)}
\chi_{-1}(x,y,t) = -\frac{2i}{\pi}\int\int\limits_C d\lambda_R d\lambda_I
\int\int\limits_C  \chi(\mu,\overline{\mu})R_0(\mu,\lambda)e^{F(\mu)-F(\lambda)}d\mu_R  d\mu_I.
\end{equation}

For delta-form kernels of the type
\begin{equation}\label{kernel1}
R_0(\mu,\overline{\mu};\lambda,\overline{\lambda})
=\sum\limits_k^{N} A_k\delta(\mu-\mu_k)\delta(\lambda-\lambda_k):=R_0(\mu,\lambda),
\end{equation}
with complex amplitudes $A_k$ and complex \,"spectral"\, points $\mu_k$, $\lambda_k$, the wave function $\chi(\lambda,\overline\lambda)$ due to (\ref{di_problem1}) has the form
 \begin{equation}\label{chi(lambda)}
\chi(\lambda,\overline\lambda)=1-\frac{2i}{\pi}\sum^N_{k=1}\frac{A_k}{\lambda-\mu_k}\chi(\mu_k)e^{F(\mu_k)-F(\lambda_k)}
\end{equation}
of the sum of $N$ terms with simple poles at \,"spectral"\, points $\lambda_k$. Such pole structure of wave function $\chi(\lambda,\overline\lambda)$ on spectral variable $\lambda$ is typical for quantum mechanics with basic Schr\"{o}dinger equation where corresponding pole structures of quantum-mechanical wave functions from wave number, energy, momentum and so on as spectral variables are commonly used. Formula (\ref{chi(lambda)})  expresses the general wave function $\chi(\lambda,\overline\lambda)$ of arbitrary spectral variables $\lambda$, $\overline\lambda$ in terms of some kind of \,"basis"\, or basic set of $N$ wave functions $\chi(\mu_k)=\chi(\mu_k,\overline\mu_k)$ at discrete $k=1,\ldots,N$ \,"spectral"\, points $\mu_k$, corresponding to the choice (\ref{kernel1}) of the kernel $R_0(\mu,\lambda)$.

From (\ref{F(lambda)}), (\ref{di_problem1}) and (\ref{kernel1}) one can obtain linear algebraic system of equations for the $N$ wave functions $\chi(\mu_k)$:
\begin{equation}\label{tildeA}
\sum\limits_{l=1}^N\tilde{A}_{kl}\chi(\mu_l)=1;\quad \tilde{A}_{kl}:=\delta_{kl}+\frac{2iA_l}{\pi(\mu_k-\lambda_l)}e^{F(\mu_l)-F(\lambda_l)}
\end{equation}
with $F(\mu)$ given by (\ref{F(lambda)}). From (\ref{F(lambda)})-(\ref{reconstruct}) one can obtain in the\, "limit of weak fields"\, restrictions on the kernel $R_0(\mu,\lambda)$ of $\overline\partial$- equation (\ref{di_problem1}) from reality condition $\overline u=u$ for KP equation (\ref{KP}):
\begin{equation}\label{RealityWeakFieldKP1}
R_0(\mu,\overline\mu;\lambda,\overline\lambda)=\overline{R_0(\overline\lambda,\lambda;\overline\mu,\mu)} - $for KP-1$,
\end{equation}
\begin{equation}\label{RealityWeakFieldKP2}
R_0(\mu,\overline\mu;\lambda,\overline\lambda)=\overline{R_0(-\overline\mu,-\mu;-\overline\lambda,-\lambda)} - $for KP-2$.
\end{equation}

From integrable boundary condition (\ref{BoundaryCondition}) we can obtain the restriction on the kernel $R_0(\mu,\lambda)$ of $\overline\partial$-equaton (\ref{di_problem1}). From (\ref{F(lambda)})-(\ref{chi_(-1)}) we derived for KP-equation (\ref{KP}):
\begin{eqnarray}\label{BoundaryConditionWeakField1}
\fl u_y\big|_{y=0}=-2i\frac{\partial^2\chi_{-1}}{\partial x\partial y}\big|_{y=0}\cong\nonumber\\
\fl-\frac{4i}{\sigma\pi}\int\int\limits_C d\lambda_{R}d\lambda_{I}\int\int\limits_C(\mu-\lambda)(\mu^2-\lambda^2)
R_0(\mu,\overline\mu;\lambda,\overline\lambda)e^{F(\mu)-F(\lambda)}\big|_{y=0}d\mu_{R}d\mu_{I}
\end{eqnarray}
The expression (\ref{BoundaryConditionWeakField1}) in the \, "limit of weak fields"\,  can be obtained as follows. For the wave function $\chi(\mu,\overline\mu)$ in integrand (\ref{BoundaryConditionWeakField1}) the approximate value $\chi(\mu,\overline\mu)\cong1$ as the first iteration $\chi(\mu,\overline\mu)\cong1$ for $\chi$ in (\ref{di_problem1}) is chosen. Due to (\ref{F(lambda)}) the phase $F(\mu)-F(\lambda)$ in exponent  (\ref{BoundaryConditionWeakField1}) has important property:
\begin{equation}\label{FF}
\left(F(-\lambda)-F(-\mu)\right)\big|_{y=0}=
\left(F(\mu)-F(\lambda)\right)\big|_{y=0},
\end{equation}
i.e. does not change at $y=0$ under change of variables $\lambda\leftrightarrow-\mu$. By the change of variables $\mu\leftrightarrow-\lambda$ in integrals (\ref{BoundaryConditionWeakField1}) we derived due to (\ref{FF})
\begin{equation}\label{BoundaryConditionWeakField2}
\fl u_y\big|_{y=0}\cong\frac{4i}{\sigma\pi}\int\int\limits_C d\mu_{R}
d\mu_{I}\int\int\limits_C(-\lambda+\mu)(\mu^2-\lambda^2)
R_0(-\lambda,-\overline\lambda,-\mu,-\overline\mu)e^{F(\mu)-F(\lambda)}\big|_{y=0}d\lambda_{R}d\lambda_{I}.
\end{equation}
Under the requirement
\begin{equation}\label{BoundaryConditionWeakField}
R_0(\mu,\overline\mu;\lambda,\overline\lambda)=R_0(-\lambda,-\overline\lambda;-\mu,-\overline\mu)
\end{equation}
from the right sides of (\ref{BoundaryConditionWeakField1}) and (\ref{BoundaryConditionWeakField2}) we concluded
\begin{equation}
u_y\big|_{y=0, (\ref{BoundaryConditionWeakField1}) }=-u_y\big|_{y=0, (\ref{BoundaryConditionWeakField2}) }\Rightarrow
u_y\big|_{y=0}=0,
\end{equation}
i.e. the restriction (\ref{BoundaryConditionWeakField}) obtained in the limit of weak fields leads to satisfaction of boundary condition (\ref{BoundaryCondition}).
It is evident that to the restriction (\ref{BoundaryConditionWeakField}) from boundary condition (\ref{BoundaryCondition}) satisfies the following delta-form kernel
\begin{equation}\label{kernel&BoundaryCond}
\fl R_0(\mu,\overline\mu;\lambda,\overline\lambda)=\sum\limits_{k=1}^N\left(a_{1k}\delta(\mu-\mu_k)\delta(\lambda-\lambda_k)+a_{1k}\delta(\mu+\lambda_k)\delta(\lambda+\mu_k)\right)
\end{equation}
with $N$ paired terms of the type $a_{1k}\delta(\mu-\mu_k)\delta(\lambda-\lambda_k)$ and $a_{1k}\delta(\mu+\lambda_k)\delta(\lambda+\mu_k)$.

All restrictions (\ref{RealityWeakFieldKP1}), (\ref{RealityWeakFieldKP2})~\cite{KonopelchenkoBook1},~\cite{KonopelchenkoBook2} (from reality condition $\overline{u}=u$) and (\ref{BoundaryConditionWeakField}) (from boundary condition (\ref{BoundaryCondition})) are obtained non rigously in the\,"limit of weak fields": for example under integrands in (\ref{BoundaryConditionWeakField1}) and (\ref{BoundaryConditionWeakField2}) the first iteration $\chi\cong1$ from $\overline\partial$-equation (\ref{F(lambda)}) is chosen. Nevertheless these restrictions (\ref{RealityWeakFieldKP1}), (\ref{RealityWeakFieldKP2}) and (\ref{BoundaryConditionWeakField}) can be used for the choice of kernel $R_0$ of $\overline\partial$-equation (\ref{di_problem1}) in constructions of exact solutions of KP equation (\ref{KP}) via general determinant form. Determinant formulas for  exact solutions of KP  can be derived by the following well known way, we  repeated derivation introducing  useful notations and rewrote corresponding formulas in convenient form.

Using general delta-form kernel $R_0$ (\ref{kernel1}) we derived the solution $\chi(\mu_k):=\chi(\mu_k,\overline\mu_k)$ $(k=1,\ldots,N)$ of the system (\ref{tildeA})
\begin{equation}\label{chi(mu_k)}
\chi(\mu_k)=\sum\limits_{l=1}^N \tilde{A}^{-1}_{kl}
\end{equation}
with matrix $\tilde{A}$ given by definition in (\ref{tildeA}). Coefficient $\chi_{-1}$ (\ref{chi_(-1)}) due to (\ref{kernel1}) and (\ref{chi(mu_k)}) took the form:
\begin{equation}\label{chi_{-1}tildeA}
\chi_{-1}=-\frac{2i}{\pi}\sum\limits_{k,l=1}^N A_k e^{F(\mu_k)-F(\lambda_k)}\tilde{A}_{kl}^{-1}=i\sum\limits_{k,l=1}^N A_{lk,x}A_{kl}^{-1},
\end{equation}
here instead of matrix $\tilde{A}$ more convenient matrix $A$ is defined by the following similarity transformation:
\begin{equation}\label{A&tildeA}
\fl A_{kl}:=e^{F(\mu_k)}\tilde{A}_{kl}e^{-F(\mu_l)}=\delta_{kl}+\frac{2iA_l}{\pi(\mu_k-\lambda_l)}e^{F(\mu_k)-F(\lambda_l)},\quad A_{kl,x}=-\frac{2A_l}{\pi}e^{F(\mu_k)-F(\lambda_l)}.
\end{equation}
Finally reconstruction formula (\ref{reconstruct}) via (\ref{chi_{-1}tildeA}), (\ref{A&tildeA}) leads to general determinant formula of exact solutions of KP equation (\ref{KP}):
\begin{equation}\label{ReconstructFinal}
u(x,y,t)=-2i\frac{\partial}{\partial x}\chi_{-1}=2\frac{\partial}{\partial x}\textrm{tr}\left(A^{-1}\frac{\partial A}{\partial x}\right)=2\frac{\partial^2}{\partial x^2}\ln\det A.
\end{equation}

Factoring from matrix $A$ multipliers $\textrm{exp}\left(\frac{1}{2}F(\mu_k)\right)$ and $\frac{2A_l}{\pi}\textrm{exp}\left(-\frac{1}{2}F(\lambda_l)\right)$ we introduced yet another more symmetrical then $A$ matrix $N$:
\begin{eqnarray}\label{MatrixN}
\fl N_{kl}:=\frac{\pi}{2A_l}e^{-\frac{1}{2}F(\mu_k)}A_{kl}e^{\frac{1}{2}F(\lambda_l)}=\frac{\pi}{2A_k}e^{-\frac{F(\mu_k)-F(\lambda_k)}{2}}\delta_{kl}+\frac{i}{\mu_k-\lambda_l}
e^{\frac{F(\mu_k)-F(\lambda_l)}{2}}=\\
\fl =\left(\frac{\pi}{2A_k}e^{-\frac{F(\mu_k)-F(\lambda_k)}{2}}+\frac{i}{\mu_k-\lambda_k}
e^{\frac{F(\mu_k)-F(\lambda_k)}{2}}\right)\delta_{kl}+\frac{i(1-\delta_{kl})}{\mu_k-\lambda_l}
e^{\frac{F(\mu_k)-F(\lambda_l)}{2}}.\nonumber
\end{eqnarray}
Using (\ref{ReconstructFinal}) and (\ref{MatrixN}) we rewrote reconstruction formula (\ref{ReconstructFinal}) in terms of matrix $N$:
\begin{equation}\label{ReconstructFinal1}
u(x,y,t)=2\frac{\partial^2}{\partial x^2}\ln\det A\equiv2\frac{\partial^2}{\partial x^2}\ln\det N.
\end{equation}

Keeping in mind the applications of determinant formula (\ref{ReconstructFinal1}) for construction in the next sections of exact multi-soliton solutions of KP-equations (\ref{KP}) with integrable boundary (\ref{BoundaryCondition}) it is convenient to specialize general formulas (\ref{MatrixN}), (\ref{ReconstructFinal1}) for the use of kernels $R_0$ of the type (\ref{kernel&BoundaryCond}). Introducing the set $A_k$ of amplitudes and the sets $M_k$, $\Lambda_k$ of spectral points
\begin{eqnarray}\label{SetsOfParameters}
&A_k:\,(a_1,a_1;a_2,a_2;\ldots;a_N,a_N),\quad (k=1,2,\ldots,2N),\\ \nonumber
&M_k:\,(\mu_1,-\lambda_1;\mu_2,-\lambda_2;\ldots,\mu_N,-\lambda_N),\\ \nonumber
&\Lambda_k:\,(\lambda_1,-\mu_1;\lambda_2,-\mu_2;\ldots,\lambda_N,-\mu_N),
\nonumber
\end{eqnarray}
we rewrote the kernel $R_0$ (\ref{kernel&BoundaryCond}) with paired terms in the form of general kernel $R_0$:
\begin{equation}\label{kernel2}
R_0(\mu,\overline\mu;\lambda,\overline\lambda)=\sum\limits_{k=1}^{2N} A_k\delta(\mu-M_k)\delta(\lambda-\Lambda_k);
\end{equation}
then general determinant formula (\ref{ReconstructFinal1}) with matrix $N_{kl}$, $(k,l=1,\ldots,2N)$ of the form
\begin{equation}\label{MatrixN1}
\fl N_{kl}:=\left(\frac{\pi}{2A_k}e^{-\frac{F(M_k)-F(\Lambda_k)}{2}}+\frac{i}{M_k-\Lambda_k}e^{\frac{F(M_k)-F(\Lambda_k)}{2}}\right)\delta_{kl}+
\frac{i(1-\delta_{kl})}{M_k-\Lambda_l}e^{\frac{F(M_k)-F(\Lambda_l)}{2}}
\end{equation}
will give exact multi-soliton solution of KP-equation with integrable boundary (\ref{BoundaryCondition}). These solutions without imposing of reality condition $\overline u=u$ are complex in general. Applying to (\ref{kernel&BoundaryCond}),  (\ref{MatrixN1}) also the restrictions from reality condition in the form (\ref{RealityWeakFieldKP1}), (\ref{RealityWeakFieldKP2}) (when (\ref{RealityWeakFieldKP1}) and (\ref{RealityWeakFieldKP2}) are working) or directly satisfying $\overline u=u$ by appropriate choice of parameters $A_k$, $M_k$, $\Lambda_k$ in determinant formulas (\ref{ReconstructFinal1})  with (\ref{kernel2}), (\ref{MatrixN1}) we derived real exact multi-soliton solutions of KP equation (\ref{KP}) with integrable boundary (\ref{BoundaryCondition}). In  the next sections 3-8  we demonstrated how general formulas (\ref{RealityWeakFieldKP1}), (\ref{RealityWeakFieldKP2}) and (\ref{chi(mu_k)})-(\ref{MatrixN1}) are working.

\section{Multi-soliton solutions of KP-1 equation, periodical on $y$}
\label{Section_3}
\setcounter{equation}{0}
\setcounter{figure}{0}
First we calculated the exact multi-soliton solutions of KP-1 equation for the case of solutions, periodical on $y$ variable and decaying on $x$, $t$ variables. Due to $F(\mu)$ (\ref{F(lambda)}) pure imaginary spectral points $\mu_k=i\mu_{k0}$, $\lambda_k=i\lambda_{k0}$ with $\overline{\mu}_{k0}=\mu_{k0}$,  $\overline{\lambda}_{k0}=\lambda_{k0}$  correspond to such solutions. Indeed in this case we derived from definition (\ref{F(lambda)}) that
\begin{equation}\label{FPeriodicY}
F(i\mu_{k0})=-\mu_{k0}x+4t\mu^3_{k0}+i\mu^2_{k0}y=-\overline{F(-i\mu_{k0})}.
\end{equation}
 Further, the phases $F(i\mu_{k0})$, which are real on $x$, $t$ variables and pure imaginary on $y$ variable, will lead to oscillating in $y$ and decaying $x$, $t$ exact solutions of KP-1 equation.
The simplest one-soliton solution corresponds to the kernel:
\begin{equation}\label{Kernel6.5R01}
R_{01}(\mu,\overline\mu;\lambda,\overline\lambda)=a_1\delta(\mu-i\mu_{10})\delta(\lambda-i\lambda_{10}),\quad \overline{\mu}_{10}=\mu_{10},\,\overline{\lambda}_{10}=\lambda_{10}.
\end{equation}
This kernel $R_{01}$ (\ref{Kernel6.5R01}) does not satisfy to restriction (\ref{RealityWeakFieldKP1}) from reality $\overline{u}=u$, because
\begin{equation}\label{Kernel6.5R02}
\fl R_{02}(\mu,\overline\mu;\lambda,\overline\lambda):=\overline{R_{01}(\overline\lambda,\lambda,\overline\mu,\mu)}=
\overline{a}_1\delta(\mu+i\lambda_{10})\delta(\lambda+i\mu_{10})\neq R_{01}(\mu,\overline\mu;\lambda,\overline\lambda).
\end{equation}

The kernels $R_{01}$ (\ref{Kernel6.5R01}) and defined by (\ref{Kernel6.5R02}) $R_{02}$  correspond to complex one-soliton solutions $u_1$ and $u_2$ which can be calculated by the use of general determinant formula (\ref{ReconstructFinal1}). For the kernel $R_{01}$ (\ref{Kernel6.5R01}) matrix $N$ (\ref{MatrixN1}) has the form
\begin{equation}\label{N11_PeriodicalY}
N_{11}=\det{N}=\frac{\pi}{2a_1}e^{-\frac{X+iY}{2}}+
\frac{1}{\mu_{10}-\lambda_{10}}e^{\frac{X+iY}{2}},
\end{equation}
here
\begin{equation}\label{DeltaFPeriodicY}
\fl F(i\mu_{10})-F(i\lambda_{10})=(\lambda_{10}-\mu_{10})x-
4t(\lambda^3_{10}-\mu^3_{10})+i(\mu^2_{10}-\lambda^2_{10})y:=X(x,t)+iY(y).
\end{equation}
Requiring
\begin{equation}
\frac{\pi}{2a_1}=\frac{1}{\mu_{10}-\lambda_{10}},\quad \overline{a}_1=a_1,
\end{equation}
we derived from (\ref{N11_PeriodicalY}) and (\ref{ReconstructFinal1}) corresponding to kernel $R_{01}$
complex one-soliton exact solution:
\begin{equation}\label{OneSolitonKP1_1}
u_1\big|_{(i\mu_{10},i\lambda_{10})}=\frac{(\lambda_{10}-\mu_{10})^2}{2\cosh^2\left(\frac{X+iY}{2}\right)}.
\end{equation}
Complex one-soliton solution $u_2(x,y,t)$ corresponding to $R_{02}$ (\ref{Kernel6.5R02}) can be calculated analogously and has the form:
\begin{equation}\label{OneSolitonKP1_(2)}
u_2\big|_{(-i\lambda_{10},-i\mu_{10})}=\frac{(\lambda_{10}-
\mu_{10})^2}{2\cosh^2\left(\frac{X-iY}{2}\right)}=
\overline{u_1\big|_{(i\mu_{10},i\lambda_{10})}},
\end{equation}
where $u_2$ can be obtained from (\ref{OneSolitonKP1_1}) by the change $\lambda_{10}\rightarrow-\mu_{10}$, $Y\rightarrow-Y$.

The general kernel $R_0$ with real amplitudes $a_k$ at $N$ paired terms of the types (\ref{Kernel6.5R01}) and (\ref{Kernel6.5R02}), i.e. the kernel
\begin{equation}\label{Kernel(KP1)1}
\fl R_0(\mu,\overline\mu;\lambda,\overline\lambda)=\sum\limits_{k=1}^N\left(a_k\delta(\mu-i\mu_{k0})\delta(\lambda-i\lambda_{k0})+
a_k\delta(\mu+i\lambda_{k0})\delta(\lambda+i\mu_{k0})\right)
\end{equation}
satisfies to both restrictions: from reality $\overline u=u$ (\ref{RealityWeakFieldKP1}) and boundary $u_y\big|_{y=0}=0$ conditions (\ref{BoundaryConditionWeakField}). This kernel with the sets $M_k$, $\Lambda_k$ of spectral points and real amplitudes $\overline a_l=a_l$, $(l=1,\ldots,N)$ (\ref{SetsOfParameters})-(\ref{MatrixN1}):
\begin{eqnarray}\label{setOfAML}
M_k=(i\mu_{10},-i\lambda_{10};i\mu_{20},-i\lambda_{20};
\ldots;i\mu_{N0},-i\lambda_{N0}),\nonumber\\
\Lambda_k=(i\lambda_{10},-i\mu_{10};i\lambda_{20},-i\mu_{20};\ldots;i\lambda_{N0},-i\mu_{N0}),
\end{eqnarray}
here $k=1,\ldots,2N$, via general determinant formulas (\ref{ReconstructFinal1}) gives exact real $2N$-soliton solution with integrable boundary condition (\ref{BoundaryCondition}).

For the simplest ($N=1$) two-soliton solution corresponding to the kernel (\ref{kernel2}) or (\ref{Kernel(KP1)1}) with $N=1$ pair of terms matrix $N$, due to (\ref{SetsOfParameters})-(\ref{MatrixN1}) and (\ref{FPeriodicY}), (\ref{DeltaFPeriodicY}), (\ref{Kernel(KP1)1}), (\ref{setOfAML}), has the form:
\begin{equation}\label{MatrixN(KP1)1}
\fl N=\left(
  \begin{array}{cc}
    \frac{\pi}{2a_1}e^{-\frac{X+iY}{2}}+\frac{1}{\mu_{10}-\lambda_{10}}e^{\frac{X+iY}{2}} & \frac{1}{2\mu_{10}}e^{\frac{F(i\mu_{10})-F(-i\mu_{10})}{2}} \\
     -\frac{1}{2\lambda_{10}}e^{\frac{F(-i\lambda_{10})-F(i\lambda_{10})}{2}}   &  \frac{\pi}{2a_1}e^{-\frac{X-iY}{2}}+\frac{1}{\mu_{10}-
     \lambda_{10}}e^{\frac{X-iY}{2}} \\
  \end{array}
\right),
\end{equation}
here
\begin{eqnarray}\label{DeltaF(KP1)1}
\fl F(i\mu_{10})-F(i\lambda_{10})=X(x,t)+iY(y),\quad F(-i\lambda_{10})-F(-i\mu_{10})=X(x,t)-iY(y);\nonumber \\
\fl F(-i\lambda_{10})-F(i\lambda_{10})=2\lambda_{10}x-8t\lambda^3_{10};\quad F(i\mu_{10})-F(-i\mu_{10})=-2\mu_{10} x+8t\mu^3_{10}.
\end{eqnarray}
For $\det N$ from (\ref{MatrixN(KP1)1}), (\ref{DeltaF(KP1)1}) we obtained the expression:
\begin{equation}\label{detN(KP1)0.5}
\det{N}=\frac{\pi^2}{4a_1^2}e^{-X}+\frac{(\mu_{10}+\lambda_{10})^2}{4\lambda_{10}\mu_{10}(\mu_{10}-\lambda_{10})^2}e^{X}+
\frac{\pi}{a_1(\mu_{10}-\lambda_{10})}\cos Y(y),
\end{equation}
here
\begin{equation}
X(x,t)=(\lambda_{10}-\mu_{10})x-4t(\lambda_{10}^3-\mu_{10}^3),\, Y(y)=(\mu^2_{10}-\lambda^2_{10})y.
\end{equation}
If we  require:
\begin{equation}\label{ConditionOnAmplitude(KP1)1}
\fl\frac{\pi^2}{a_1^2}=\frac{(\mu_{10}+\lambda_{10})^2}{\lambda_{10}\mu_{10}(\mu_{10}-\lambda_{10})^2}\Rightarrow
\frac{\pi}{a_1}=\pm\frac{\mu_{10}+\lambda_{10}}{\sqrt{\lambda_{10}
\mu_{10}}(\mu_{10}-\lambda_{10})},\quad \lambda_{10}\mu_{10}>0,
\end{equation}
then due to (\ref{detN(KP1)0.5})
\begin{equation}\label{detN(KP1)1}
\fl \det{N}=\frac{(\mu_{10}+\lambda_{10})^2}{2\lambda_{10}\mu_{10}(\mu_{10}-\lambda_{10})^2}\left(\cosh X(x,t)+\alpha\cos Y(y)\right)>0,\quad \alpha:=\pm\frac{2\sqrt{\lambda_{10}\mu_{10}}}{\mu_{10}+\lambda_{10}}.
\end{equation}

The reconstruction formula (\ref{ReconstructFinal1}) with (\ref{ConditionOnAmplitude(KP1)1}), (\ref{detN(KP1)1}) gives the exact real nonsingular, due to $|\alpha|<1$ (\ref{detN(KP1)1}), two-soliton solution of KP-1 equation (\ref{KP}) with integrable boundary (\ref{BoundaryCondition}):
\begin{equation}\label{TwoSolitonKP1(1)}
u(x,y,t)=2\frac{(\lambda_{10}-\mu_{10})^2(1+\alpha \cosh X(x,t)\cos Y(y))}{(\cosh X(x,t)+\alpha \cos Y(y))^2}.
\end{equation}
\begin{figure}[h]
\begin{center}
\includegraphics[width=0.50\textwidth, keepaspectratio]{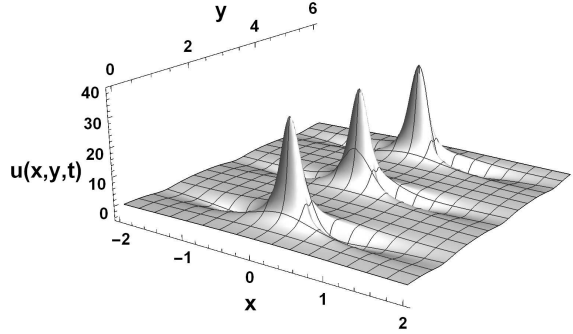}
\parbox[t]{1\textwidth}{\caption{Two-soliton, periodical on $y$, solution of KP-1 $u$ (\ref{TwoSolitonKP1(1)})  with parameters $\lambda_{10}=1$, $\mu_{10}=2$.}\label{pictPeriodicOnYKP1}}
\end{center}
\end{figure}
This exact nonsingular, due to $|\alpha|<1$ (\ref{detN(KP1)1}), one-periodical on the variable $Y(y)$ solution,  
is even function of $y$, therefore $u_y\big|_{y=0}=0$, and integrable boundary condition (\ref{BoundaryCondition}) is satisfied.
The graph of this solution on semi-plane $y\geq0$ is shown on figure  (\ref{pictPeriodicOnYKP1}).
Introducing modified or deformed phase $Y_D(y)$ by definition
\begin{equation}
\cos Y_D(y):=\alpha\cos Y(y)
\end{equation}
and using the identity
\begin{equation}\label{Identity}
\frac{1+\cosh X\cos Y_D}{(\cosh X+\cos Y_D)^2}=\frac{1}{4}\left(\frac{1}{\cosh^2\frac{X+iY_D}{2}}+
\frac{1}{\cosh^2\frac{X-iY_D}{2}}\right),
\end{equation}
we rewrote the exact solution (\ref{TwoSolitonKP1(1)}) of KP-1 as nonlinear superposition of exact complex solutions (\ref{OneSolitonKP1_1})  and (\ref{OneSolitonKP1_(2)}) with $Y(y)$ changed by $Y_D(y)$:
\begin{eqnarray}\label{Superposition1_KP1}
u(x,y,t)=u_1\big|_{(i\mu_{10},\:i\lambda_{10})\, Y\rightarrow Y_D}+u_2\big|_{(-i\lambda_{10},\:-i\mu_0)\, Y\rightarrow Y_D}=\nonumber\\
=\frac{(\lambda_{10}-\mu_{10})^2}{2}\left(\frac{1}{\cosh^2\frac{X+iY_D}{2}}
+\frac{1}{\cosh^2\frac{X-iY_D}{2}}\right).
\end{eqnarray}
Two terms in (\ref{Superposition1_KP1}) did not coincide with $u_1$ (\ref{OneSolitonKP1_1}) and  $u_2$ (\ref{OneSolitonKP1_(2)}) (with $Y(y)$ changed by $Y_D(y)$), these terms are not the exact solutions of KP-1, but their sum is exact real two-soliton solution of KP-1 equation (\ref{KP}).
 The exact solution (\ref{TwoSolitonKP1(1)}) or (\ref{Superposition1_KP1}) corresponds to "bound state" of two complex one-solitons (\ref{OneSolitonKP1_1})  and (\ref{OneSolitonKP1_(2)}); this bound state arises due to imposition of boundary condition (\ref{BoundaryCondition}) and represents certain eigen-mode oscillations of the field $u(x,y,t)$ in semi-plane $y\geq0$; this resembles the formation of standing wave on elastic string due to corresponding boundary conditions at endpoints of the string.

\section{Multi-soliton solutions of KP-1 equation, pure solitonic case}
\label{Section_4}
\setcounter{equation}{0}
The kernels $R_0$ with conjugate to each other $\overline\mu_k=\lambda_k$ spectral points correspond to pure solitonic solutions of KP-1 equation with nonoscillating behaviour on all coordinates $x$, $y$ and time $t$ variables. Indeed, for such kernels the phases $F(\mu_k)-F(\overline\mu_k)$ due to (\ref{F(lambda)})
\begin{eqnarray}\label{XY}
&F(\mu_k)-F(\overline\mu_k)=i(\mu_k-\overline\mu_k)x-i
(\mu^2_k-\overline\mu^2_k)y+4it(\mu^3_k-\overline\mu^3_k)=\nonumber\\
&-2\mu_{kI}x+4\mu_{kR}\mu_{kI}y+8t(\mu^3_{kI}-3\mu_{kR}^2\mu_{kI}):=
-2\left(X_k(x,t)+Y_k(y)\right)
\end{eqnarray}
are real and nonoscillatory and one should to expect that corresponding multi-solitons solutions will be nonperiodical on all space-time variables.

The simplest one-soliton solution corresponds to the kernel:
\begin{equation}\label{Kernel7R01}
R_{01}(\mu,\overline\mu;\lambda,\overline\lambda)=
a_1\delta(\mu-\mu_1)\delta(\lambda-\overline\mu_1).
\end{equation}
This kernel $R_{01}$ due to the relation
\begin{equation}\label{overlineKernel7R01}
\overline{R_{01}(\overline\lambda,\lambda,\overline\mu,\mu)}=\overline{a}_1\delta(\mu-\mu_1)\delta(\lambda-\overline\mu_1)=R_{01}(\mu,\overline\mu;\lambda,\overline\lambda)
\end{equation}
for $\overline{a}_1=a_1$ satisfies the restriction of the reality condition (\ref{RealityWeakFieldKP1}) and  via general formulas (\ref{A&tildeA}), (\ref{ReconstructFinal}) leads to corresponding exact real one-soliton solution. In considered case matrix $N$ (\ref{MatrixN1}) has the form:
\begin{equation}\label{N11_PureSolitonic}
N_{11}=\det{N}=\frac{\pi e^{X-Y}}{2a_1}+\frac{e^{-(X-Y)}}{2\mu_{1I}},
\end{equation}
here due to (\ref{XY})
\begin{equation}
X(x,t):=\mu_{1I}x-4t(\mu^3_{1I}-3\mu^2_{1R}\mu_{1I}),\quad Y(y):=-2\mu_{1R}\mu_{1I}y.
\end{equation}
Requiring
\begin{equation}
\frac{\pi}{a_1}=\frac{1}{\mu_{1I}},
\end{equation}
using  reconstruction formula (\ref{ReconstructFinal1}) we derived exact real nonsingular one-soliton solution $u_1(x,y,t)$ of KP-1 equation (\ref{KP}):
\begin{equation}\label{u1OneSolitonKP1_2}
u_1(x,y,t)\big|_{(\mu_1,\overline\mu_1)}=2\frac{\partial^2}{\partial x^2}\ln\det N=\frac{2\mu^2_{1I}}{\cosh^2(X-Y)}.
\end{equation}
Exact real nonsingular one-soliton solution $u_2(x,y,t)$  that corresponds to the kernel
\begin{equation}\label{Kernel7R02}
R_{02}(\mu,\overline\lambda;\mu,\overline\mu):=a_1\delta(\mu+\overline\mu_1)\delta(\lambda+\mu_1),\quad \overline{a}_1=a_1
\end{equation}
can be derived from (\ref{u1OneSolitonKP1_2}) by the simple changes $\mu_{1R}\rightarrow -\mu_{1R}$, $Y\rightarrow -Y$:
\begin{equation}\label{u2OneSolitonKP1_2}
u_2(x,y,t)\big|_{(-\overline\mu_1,-\mu_1)}=2\frac{\partial^2}{\partial x^2}\ln\det N=\frac{2\mu^2_{1I}}{\cosh^2(X+Y)}.
\end{equation}
The solutions $u_{1}$ and $u_{2}$ do not satisfy to boundary condition (\ref{BoundaryCondition}).

The general kernel $R_0$ of $\overline\partial$-equation (\ref{di_problem1}) with sum of pairs of the type $R_{01}$ (\ref{Kernel7R01}) and $R_{02}$ (\ref{Kernel7R02}) with real amplitudes $\overline{a}_k=a_k$, i.e. the kernel
\begin{equation}\label{Kernel(KP1)3}
 R_0(\mu,\overline\mu;\lambda,\overline\lambda)=\sum\limits_{k=1}^N\left(a_k\delta(\mu-\mu_{k})\delta(\lambda-\overline{\mu}_{k})+
a_k\delta(\mu+\overline{\mu}_{k})\delta(\lambda+\mu_{k})\right)
\end{equation}
evidently satisfies the restriction (\ref{BoundaryConditionWeakField}) from boundary condition (\ref{BoundaryCondition}) $u_y\big|_{y=0}=0$ and also the restriction from reality condition due to (\ref{Kernel7R01}),  (\ref{overlineKernel7R01}) (\ref{Kernel7R02}). Therefore this kernel leads to exact real $2N$-soliton solutions of KP-1 equation with integrable boundary (\ref{BoundaryCondition}). Introducing the set of amplitudes $A_k$ with real amplitudes $\overline{a}_k=a_k$ and sets $M_k$, $\Lambda_k$ of spectral points
\begin{eqnarray}\label{setOfAML3}
\fl A_k:\,(a_1,a_1;a_2,a_2;\ldots;a_N,a_N), \quad (k=1,...,2N);\nonumber\\
\fl M_k:\,(\mu_{1},-\overline{\mu}_1;\mu_{2},-\overline{\mu}_2;\ldots;\mu_{N},
-\overline{\mu}_N),\quad
\Lambda_k:\,(\overline{\mu}_1,-\mu_{1};\overline{\mu}_2,-\mu_{2};\ldots;
\overline{\mu}_N,-\mu_{N}),
\end{eqnarray}
we rewrote the kernel (\ref{Kernel(KP1)3}) in general form (\ref{kernel1}) with matrix $N$ of the form (\ref{MatrixN1}); finally the reconstruction formula (\ref{ReconstructFinal1}) with matrix $N$ (\ref{MatrixN1}) gives exact $2N$-soliton solution of KP-1 equation with integrable boundary (\ref{BoundaryCondition}).

The simplest two-soliton solution of considered type corresponds to one pair of terms in (\ref{Kernel(KP1)3}), matrix $N$ for this case has the form:
\begin{equation}\label{MatrixN(KP1)3}
 N=
\left(
  \begin{array}{cc}
    \frac{\pi}{2a_1}e^{X-Y}+\frac{i}{\mu_1-\overline\mu_1}e^{-(X-Y)} & \frac{i}{2\mu_1}e^{\frac{F(\mu_1)-F(-\mu_1)}{2}} \\
     -\frac{i}{2\overline\mu_1}e^{\frac{F(-\overline\mu_1)- F(\overline\mu_1)}{2}}   &  \frac{\pi}{2a_1}e^{X+Y}+\frac{i}{\mu_1-\overline\mu_1}e^{-(X+Y)} \\
  \end{array}
\right),
\end{equation}
here
\begin{eqnarray}\label{DeltaF(KP1)3}
&F(\mu_1)-F(-\mu_1) +F(-\overline\mu_1)-F(\overline\mu_1)=
 -4X(x,t) ;\nonumber\\
&X(x,t)=\mu_{1I}x-4t(\mu_{1I}^3-3\mu_{1R}^2\mu_{1I}),\, Y(y)=-2\mu_{1R}\mu_{1I}y.
\end{eqnarray}
For $\det N$ from (\ref{MatrixN(KP1)3}), (\ref{DeltaF(KP1)3}) we derived the expression:
\begin{equation}
\det{N}=\frac{\pi^2}{4a_1^2}e^{2X}+\frac{\mu_{1R}^2}{4\mu_{1I}^2|\mu_1|^2}e^{-2X}+\frac{\pi}{4a_1\mu_{1I}}e^{-2Y}+
\frac{\pi}{4a_1\mu_{1I}}e^{2Y}.
\end{equation}
If
\begin{equation}\label{ConditionOnAmplitude(KP1)3}
\frac{\pi^2}{a_1^2}=\frac{\mu_{1R}^2}{\mu_{1I}^2|\mu_{1}|^2}\Rightarrow
\frac{\pi}{a_1}=\pm\frac{\mu_{1R}}{\mu_{1I}|\mu_{1}|},
\end{equation}
then due to (\ref{ConditionOnAmplitude(KP1)3})
\begin{equation}\label{detN(KP1)3}
\det{N}=
\frac{\mu_{1R}^2}{2\mu_{1I}^2|\mu_1|^2}\left(\cosh 2X+\alpha\cosh 2Y\right), \quad \alpha=\pm\frac{|\mu_1|}{\mu_{1R}}.
\end{equation}
Reconstruction formula (\ref{ReconstructFinal1}) gives the exact real nonsingular two-soliton solution of KP-1 equation (\ref{KP}) with integrable boundary (\ref{BoundaryCondition}):
\begin{equation}\label{TwoSolitonKP1(1)3}
u(x,y,t)=2\frac{\partial^2}{\partial x^2}\ln\det N=8\mu^2_{1I}\frac{1+\alpha \cosh 2X(x,t)\cosh 2Y(y)}{(\cosh 2X(x,t)+\alpha \cosh 2Y(y))^2},
\end{equation}
for $\alpha>-1$ this pure solitonic (decaying) solution is nonsingular.
This solution indeed satisfies the condition (\ref{BoundaryCondition}) due to the fact that (\ref{TwoSolitonKP1(1)3}) is even function $y$.
The graph of this solution on semi-plane $y\geq 0$ is shown on figure (\ref{PureSolitonicKP1}).

\begin{figure}[h]\label{PureSolitonicKP1}
\begin{center}
\includegraphics[width=0.50\textwidth, keepaspectratio]{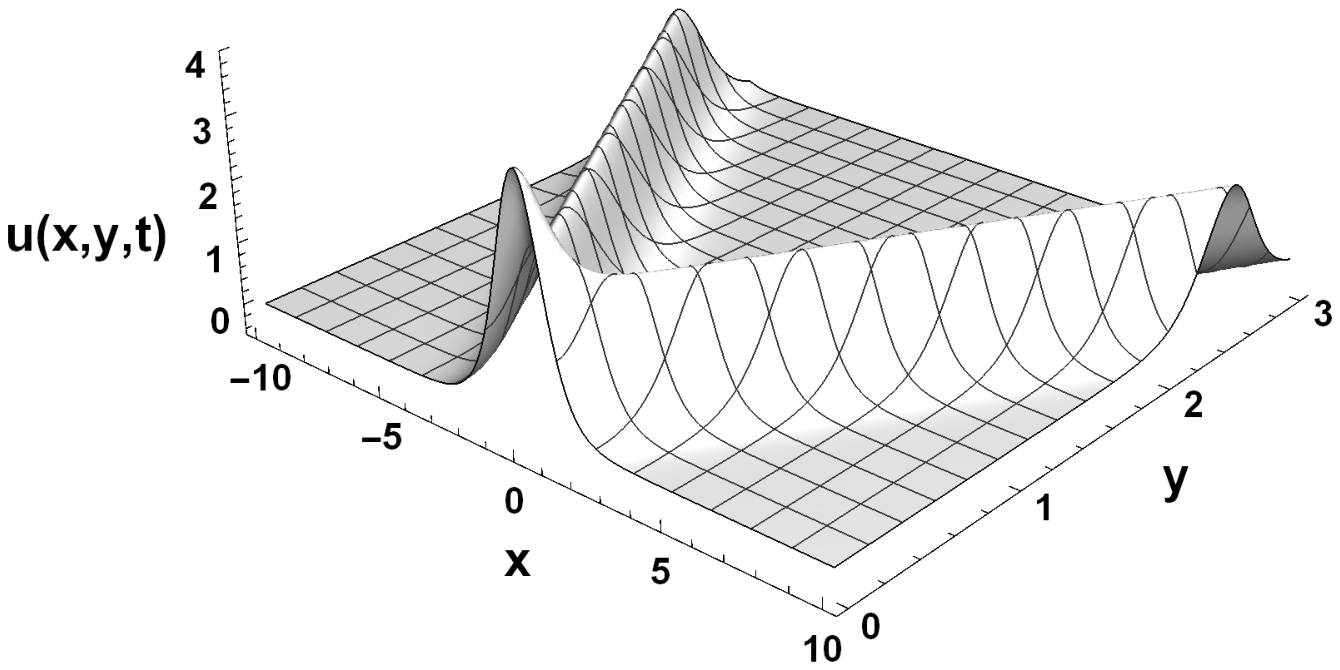}
\parbox[t]{1\textwidth}{\caption{Two-soliton solution of KP-1 $u$ (\ref{TwoSolitonKP1(1)3})  with parameters $\mu_{1I}=1$, $\mu_{1R}=2$.}\label{PureSolitonicKP1}}
\end{center}
\end{figure}

Using the identity
\begin{equation}
\frac{1+\cosh 2X\cosh 2Y}{(\cosh 2X+\cosh 2Y)^2}=\frac{1}{4}\left(\frac{1}{\cosh^2\left(X+Y\right)}+\frac{1}{\cosh^2\left(X-Y\right)}\right)
\end{equation}
and defining for $\alpha>0$ instead $Y(y)$ new phase $Y_D(y)$
\begin{equation}\label{DeformedPhaseY}
\cosh 2Y_D(y):=\alpha\cosh 2Y(y),
\end{equation}
we were able to represent the exact solution (\ref{TwoSolitonKP1(1)3}) as nonlinear superposition of exact solutions (\ref{u1OneSolitonKP1_2}) and (\ref{u2OneSolitonKP1_2}) with deformed phase $Y_D$  (\ref{DeformedPhaseY}) in the form of the following sum:
\begin{equation}\label{Superposition}
\fl u(x,y,t)=\frac{2\mu_{1I}^2}{\cosh^2(X-Y_D)}+\frac{2\mu_{1I}^2}{\cosh^2(X+Y_D)}
=u_1\big|_{(\mu_1,\:\overline\mu_1),\, Y\rightarrow Y_D}+u_2\big|_{(-\mu_1,\:-\overline\mu_1),\, Y\rightarrow Y_D},
\end{equation}
here $u_1\big|_{(\mu_1,\:\overline\mu_1),\, Y\rightarrow Y_D}$ and $u_2\big|_{(-\mu_1,\:-\overline\mu_1),\, Y\rightarrow Y_D}$ are defined by (\ref{u1OneSolitonKP1_2}) and (\ref{u2OneSolitonKP1_2}) with $Y_D$  (\ref{DeformedPhaseY}).
Two terms in (\ref{Superposition}) with $Y(y)$ changed by $Y_D(y)$ did not coincide with $u_1$ (\ref{u1OneSolitonKP1_2}) and  $u_2$ (\ref{u2OneSolitonKP1_2}), these terms are not the exact solutions of KP-1, but their sum is exact real two-soliton solution of KP-1 equation (\ref{KP}).
The exact solution (\ref{TwoSolitonKP1(1)3}) or (\ref{Superposition}) corresponds to "bound state" of two  one-solitons (\ref{u1OneSolitonKP1_2})  and (\ref{u2OneSolitonKP1_2}). The solution (\ref{Superposition}) arises due to imposition of boundary condition (\ref{BoundaryCondition}) and represents certain eigen-mode of the field $u(x,y,t)$ in semi-plane $y\geq0$ propagating along $x$-axis with velocity $V_x=4(\mu_{1I}^2-3\mu_{1R}^2)$; this resembles the formation of standing waves on elastic string due to corresponding boundary conditions at endpoints of the string.

\section{Multi-soliton solutions of KP-1 equation with integrable boundaries, periodic on $X(x,t)$ variables}
\label{Section_5}
\setcounter{equation}{0}
To multi-soliton solutions, periodical on $x$, $t$-variables, leads delta-form kernel $R_0$ of $\overline\partial$-equation (\ref{di_problem1}) with spectral points $\lambda_k=-\overline\mu_k$. For such kernel the phases $F(\mu_k)-F(-\overline\mu_k)$
\begin{eqnarray}\label{XY_KP_1_3}
F(\mu_k)-F(-\overline\mu_k)=i(\mu_k+\overline\mu_k)x-i(\mu^2_k-
\overline\mu^2_k)y+4it(\mu^3_k+\overline\mu^3_k)= \nonumber \\
i\left(2\mu_{kR}x+8t(\mu^3_{kR}-3\mu_{kR}\mu^2_{kI})\right)
+4\mu_{kR}\mu_{kI}y:=2\left(iX_k(x,t)+Y_k(y)\right)
\end{eqnarray}
are pure imaginary on $X(x,t)$ variable and real on $Y(y)$ variable; this corresponds to oscillatory behavior of solutions on $X(x,t)$ and decaying behavior on $Y_k(y)$.

The simplest one-soliton solution of considered type corresponds to the kernel $R_{01}$
\begin{equation}\label{Kernel8R01}
R_{01}(\mu,\overline\mu;\lambda,\overline\lambda)=
a_1\delta(\mu-\mu_1)\delta(\lambda+\overline\mu_1),
\end{equation}
the use of this kernel leads to the following matrix $N$ (\ref{MatrixN1}) and $\det{N}$:
\begin{eqnarray}\label{N11_PeriodicXT}
 N_{11}=\det{N}=\left(\frac{\pi}{2a_1}e^{-\frac{F(\mu_1)-
F(-\overline\mu_1)}{2}}+\frac{i}{\mu_{1}+\overline{\mu}_1}
e^{\frac{F(\mu_1)-F(-\overline\mu_1)}{2}}\right)
\bigg|_{\frac{\pi}{a_1}=\frac{1}{\mu_{1R}}}=\nonumber\\
=\frac{e^{i\frac{\pi}{4}}}{\mu_{1R}}\cosh\left(i\left(X+Y+\frac{\pi}{4}\right)
\right).
\end{eqnarray}
In (\ref{N11_PeriodicXT}) for real amplitude $a_1$ the following relation with spectral point $\mu_1=\mu_{1R}+i\mu_{1I}$ is chosen:
\begin{equation}
\frac{\pi}{a_1}=\frac{1}{\mu_{1R}}.
\end{equation}
By reconstruction formula (\ref{ReconstructFinal1}) we derived with (\ref{N11_PeriodicXT}) the exact complex-valued one-soliton solution of KP-1 equation (\ref{KP})
\begin{equation}\label{u1OneSolitonKP1_3}
\fl u_1(x,y,t)\big|_{(\mu_1,-\overline\mu_1)}=2\frac{\partial^2}{\partial x^2}\ln\det N=-\frac{2\mu^2_{1R}}{\cosh^2\left(Y+i\left(X+\frac{\pi}{4}\right)\right)}=
-\frac{2\mu^2_{1R}}{\cos^2\left(X+\frac{\pi}{4}-iY\right)}.
\end{equation}
The solution (\ref{u1OneSolitonKP1_3}) evidently periodical in $X(x,t)$ variable and decaying in $Y(y)$ variable.

In order to construct exact periodical on $X(x,t)$ solution $u(x,y,t)$ satisfying to integrable boundary condition (\ref{BoundaryCondition}) we added to kernel (\ref{Kernel8R01}) $R_{01}$  another kernel $R_{02}$ defined by the relation:
\begin{equation}\label{Kernel8R02}
R_{02}(\mu,\overline\mu;\lambda,\overline\lambda):=
R_{01}(-\lambda,-\overline\lambda;-\mu,-\overline{\lambda})=
a_1\delta(\mu-\overline{\mu_1})\delta(\lambda+\mu_1).
\end{equation}
The exact one-soliton solution $u_2(x,y,t)\big|_{(\overline\mu_1,-\mu_1)}$ corresponding to (\ref{Kernel8R02}) can be calculated analogously to $u_1(x,y,t)\big|_{(\mu_1,-\overline\mu_1)}$ and equal to $u_1$ given by (\ref{u1OneSolitonKP1_3}) with the change $\mu_{1I}\rightarrow -\mu_{1I}$ and consequently, due to definitions of $X(x,t)$ and $Y(y)$ in (\ref{XY_KP_1_3}),  by the change $Y\rightarrow -Y$ in (\ref{u1OneSolitonKP1_3}); so one obtains for $u_2$:
\begin{eqnarray}\label{u2OneSolitonKP1_3}
\fl u_2(x,y,t)\big|_{(\overline\mu_1,-\mu_1)}=
-\frac{2\mu^2_{1R}}{\cosh^2\left(-Y+i\left(X+\frac{\pi}{4}\right)\right)}=
-\frac{2\mu^2_{1R}}{\cos^2\left(X+\frac{\pi}{4}+iY\right)}=\nonumber\\
\fl =\overline{u_1(x,y,t)\big|_{(\mu_1,-\overline\mu_1)}}.
\end{eqnarray}

One-soliton solutions (\ref{u1OneSolitonKP1_3}) and (\ref{u2OneSolitonKP1_3}) do not satisfy to integrable boundary condition (\ref{BoundaryCondition}), however they are usefull for the constructions of present section.
The sum $R_{01}+R_{02}$ of considered kernels (\ref{Kernel8R01}) and (\ref{Kernel8R02}) satisfies the restriction (\ref{BoundaryConditionWeakField}) from boundary condition (\ref{BoundaryCondition}), the same restriction is  satisfied by more general kernel with $N$ pairs of terms of the type (\ref{Kernel8R01}) and (\ref{Kernel8R02}) with real amplitudes $a_k$:
\begin{equation}\label{Kernel(KP1)4}
\fl R_0(\mu,\overline\mu;\lambda,\overline\lambda)=\sum\limits_{k=1}^N\left(a_k\delta(\mu-\mu_{k})\delta(\lambda+\overline{\mu}_{k})+
a_k\delta(\mu-\overline{\mu}_{k})\delta(\lambda+\mu_{k})\right).
\end{equation}
This kernel can be rewritten in general form (\ref{kernel2}) with the set $A_k$ of amplitudes (\ref{SetsOfParameters}) and the sets $M_k$, $\Lambda_k$ of spectral points:
\begin{eqnarray}\label{setOfAML4}
\fl A_k:\,(a_1,a_1;a_2,a_2;\ldots;a_N,a_N),\quad (k=1,...,2N);\nonumber\\
\fl M_k:\,(\mu_{1},\overline{\mu}_1;\mu_{2},\overline{\mu}_2;\ldots;\mu_{N},
\overline{\mu}_N), \quad \Lambda_k:\,(-\overline{\mu}_1,-\mu_{1};-\overline{\mu}_2,-
\mu_{2};\ldots;-\overline{\mu}_N,-\mu_{N}).
\end{eqnarray}
Then general formulas (\ref{MatrixN1}), (\ref{ReconstructFinal1}) give via reconstruction formula the exact
multi-soliton solution of KP-1 (\ref{ReconstructFinal1}) in determinant form.

In the simplest case of one pair $N=1$  of terms in (\ref{kernel2})
we have for matrix $N$ (\ref{MatrixN1}):
\begin{equation}\label{MatrixN(KP1)4}
\left(
  \begin{array}{cc}
    \frac{\pi}{2a_1}e^{-iX-Y}+\frac{i}{\mu_1+\overline\mu_1}e^{iX+Y} & \frac{i}{2\mu_1}e^{\frac{F(\mu_1)-F(-\mu_1)}{2}} \\
     \frac{i}{2\overline\mu_1}e^{\frac{F(\overline\mu_1)-F(-\overline\mu_1)}{2}}   &  \frac{\pi}{2a_1}e^{-iX+Y}+\frac{i}{\mu_1+\overline\mu_1}e^{iX-Y} \\
  \end{array}
\right),
\end{equation}
here
\begin{eqnarray}\label{DeltaF(KP1)4}
\fl F(\mu_1)-F(-\mu_1) =i\left(2\mu_1x+8t\mu^3_1\right),\quad F(\overline\mu_1)-F(-\overline\mu_1) =i\left(2\overline\mu_1x+8t\overline{\mu}^3_1\right),\nonumber\\
\fl F(\mu_1)-F(-\mu_1)+F(\overline\mu_1)-F(-\overline\mu_1)=
i\left(4\mu_{1R}x+16t(\mu_{1R}^3-3\mu_{1R}\mu^2_{1I})\right),\nonumber\\
\fl X(x,t):=\mu_{1R}x+4t(\mu_{1R}^3-3\mu_{1R}\mu_{1I}^2),\, Y(y):=2\mu_{1R}\mu_{1I}y.
\end{eqnarray}
Suggesting that
\begin{equation}
\frac{\pi}{a_1}=\pm\frac{\mu_{1I}}{\mu_{1R}|\mu_1|},
\end{equation}
we derived for $\det N$ from (\ref{MatrixN(KP1)4}) and (\ref{DeltaF(KP1)4}):
 \begin{eqnarray}
&\det{N}=\frac{\pi^2}{4a_1^2}e^{-2iX}-
\frac{\mu_{1I}^2}{4\mu_{1R}^2|\mu_1|^2}e^{2iX}+\frac{\pi i}{4a_1\mu_{1R}}e^{2Y}+
\frac{\pi i}{4a_1\mu_{1R}}e^{-2Y}=\nonumber\\
&\frac{\mu^2_{1I}i}{2\mu^2_{1R}|\mu_1|^2}\left(\cos\left(2X+\frac{\pi}{2}\right)+
\left(\pm\frac{|\mu_1|}{\mu_{1I}}\right)\cosh 2Y\right).
\end{eqnarray}
Therefore reconstruction formula (\ref{ReconstructFinal1}) gives the exact real nonsingular, due to $|\alpha |>1$, two-soliton solution of KP-1 equation satisfying integrable boundary condition (\ref{BoundaryCondition}):
\begin{equation}\label{TwoSolitonKP1(1)4}
\fl u(x,y,t)=2\frac{\partial^2}{\partial x^2}\ln\det N=-8\mu^2_{1R}\frac{1+\alpha \cosh 2Y\cos\left(2X+\frac{\pi}{2}\right)}{(\cos\left(2X+
\frac{\pi}{2}\right)+\alpha \cosh 2Y)^2},\quad \alpha=\pm\frac{|\mu_1|}{\mu_{1I}},
\end{equation}
where boundary condition (\ref{BoundaryCondition}) is satisfied due to the fact that $u(x,y,t)$ (\ref{TwoSolitonKP1(1)4}) is even function of $y$ (through the $\cosh 2Y(y)$). The considered solution propagates along $x$-axis with velocity $V_x=-4(\mu^2_{1R}-3\mu^2_{1I})$.

The graph of solution (\ref{TwoSolitonKP1(1)4}) on semi-plane $y\geq 0$ is shown on figure (\ref{PeriodicOnXTKP1}).
\begin{figure}[h]\label{PeriodicOnXTKP1}
\begin{center}
\includegraphics[width=0.8\textwidth, keepaspectratio]{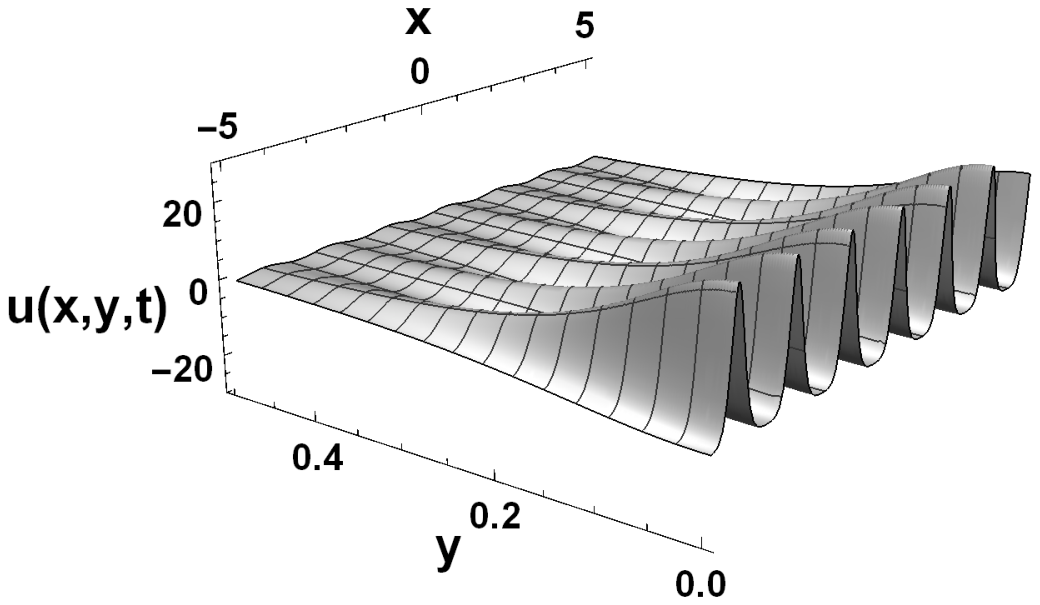}
\parbox[t]{1\textwidth}{\caption{Two-soliton solution of KP-1 $u$ (\ref{TwoSolitonKP1(1)4})  with parameter $\mu_{1I}=1$, $\mu_{1R}=2$.}\label{PeriodicOnXTKP1}}
\end{center}
\end{figure}
This solution due to identity (\ref{Identity}) can be rewritten as nonlinear superposition of two complex-valued one-solitons (\ref{u1OneSolitonKP1_3}) and (\ref{u2OneSolitonKP1_3}) in the form of the sum of one-soliton solutions (\ref{u1OneSolitonKP1_3}) and (\ref{u2OneSolitonKP1_3}) with modified phase $Y\rightarrow Y_D$:
\begin{eqnarray}\label{Superposition2}
u(x,y,t)=u_1\big|_{(\mu_1,\:-\overline\mu_1) Y\rightarrow Y_D}+u_2\big|_{(\overline\mu_1,-\mu_1)Y\rightarrow Y_D}=\nonumber\\
=-\frac{2\mu_{1R}^2}{\cos^2\left(X+\frac{\pi}{4}-iY_D\right)}-
\frac{2\mu_{1R}^2}{\cos^2\left(X+\frac{\pi}{4}+iY_D\right)},
\end{eqnarray}
here for $\alpha>0$ modified or deformed phase $Y_D$ is defined by the formula:
\begin{equation}\label{DeformedPhaseY1}
\cosh 2Y_{D}(y):=\alpha\cosh 2Y(y)
\end{equation}
Two terms in (\ref{Superposition2}) with $Y_D(y)$ do not coincide with $u_1$ (\ref{u1OneSolitonKP1_3}) and  $u_2$ (\ref{u2OneSolitonKP1_3}). These terms are not the exact solutions of KP-1, but their sum is exact real two-soliton solution of KP-1 equation (\ref{KP}).
The exact two-soliton solution (\ref{TwoSolitonKP1(1)4}) represents \,"bound state"\, of two simple complex-valued one-soliton solutions (\ref{u1OneSolitonKP1_3}) and (\ref{u2OneSolitonKP1_3}), this solution
resembles certain eigenmode of oscillations of field $u(x,y,t)$ in semi-plane $y\geq 0$ arising as result of imposition of boundary condition (\ref{BoundaryCondition}) on this field.

\section{Multi-soliton solutions of KP-2 equation with integrable boundary, periodical on $y$ variable}
\label{Section_6}
\setcounter{equation}{0}
One-soliton exact solution of KP-2 equation, periodical on $y$, can be constructed by the use of the following $R_{01}$ kernel:
\begin{equation}\label{R02KP2_1}
R_{01}(\mu,\overline\mu;\lambda,\overline\lambda)=
a_1\delta(\mu-\mu_1)\delta(\lambda-\overline\mu_1).
\end{equation}
The phase $F(\mu_1)-F(\overline{\mu}_1)$ for this kernel, due to (\ref{F(lambda)}) with $\sigma=1$, has the form:
\begin{eqnarray}
F(\mu_1)-F(\overline\mu_1)=i(\mu_1-\overline\mu_1)x+
(\mu^2_1-\overline\mu^2_1)y+4it(\mu^3_1-\overline\mu^3_1)=\nonumber\\
=-2\mu_{1I}x+8t\left(\mu^3_{1I}-3\mu_{1I}\mu^2_{1R})\right)
+4i\mu_{1R}\mu_{1I}y:=-2X(x,t)+i2Y(y),\nonumber\\
X(t):=\mu_{1I}x-4t\left(\mu^3_{1I}-3\mu_{1I}\mu^2_{1R})\right)\,, \quad Y(y):=2\mu_{1R}\mu_{1I}y.
\end{eqnarray}
Matrix $N_{kl}$ and $\det N$ in this case are given by expression
\begin{equation}\label{N11_PeriodicY_KP2}
N_{11}=\det{N}=\frac{\pi}{2a_1}e^{X-iY}+\frac{1}{2\mu_{1I}}e^{-X+iY}
=\frac{1}{\mu_{1I}}\cosh\left(X-iY\right).
\end{equation}
In (\ref{N11_PeriodicY_KP2}) real amplitude $a_1$ connecting with with spectral point $\mu_1=\mu_{1R}+i\mu_{1I}$  by following relation
\begin{equation}
\frac{\pi}{a_1}=\frac{1}{\mu_{1I}}
\end{equation}
is chosen.

Reconstruction formula (\ref{ReconstructFinal1}) leads due to (\ref{N11_PeriodicY_KP2}) to complex exact solution of KP-2:
\begin{equation}\label{u1OneSolitonKP2_11}
u_1(x,y,t)\big|_{(\mu_1,\overline\mu_1)}=2\frac{\partial^2}{\partial x^2}\ln\det N
=\frac{2\mu^2_{1I}}{\cosh^2\left(X-iY\right)}
\end{equation}
For the kernel $R_{02}(\mu,\overline\mu;\lambda,\overline\lambda)$, defined through $R_{01}$  (\ref{R02KP2_1}) in accordance with  (\ref{RealityWeakFieldKP2}) by the formula:
\begin{equation}\label{Kernel8R02KP2}
\fl R_{02}(\lambda,\overline\lambda;\mu,\overline\mu):=
\overline{R_{01}(-\overline\mu,-\mu;-\overline\lambda,-\lambda)}=
\overline{a_1}\delta(\mu+\overline\mu_1)\delta(\lambda+\mu_1),\quad a_1=\overline{a_1},
\end{equation}
corresponds the phase
\begin{equation}
\fl F(-\overline\mu_1)-F(-\mu_1)=i(-\overline\mu_1+\mu_1)x+(\overline\mu^2_1-\mu^2_1)y-4it(-\overline\mu^3_1+\mu^3_1)
=-2X-i2Y
\end{equation}
and exact complex solution
\begin{equation}\label{u2OneSolitonKP2_11}
u_2(x,y,t)\big|_{(-\overline\mu_1,-\mu_1)}=
\frac{2\mu^2_{1I}}{\cosh^2\left(X+iY\right)}=
\overline{{u}_1\big|_{(\mu_1,\overline\mu_1)}}.
\end{equation}

The kernel $R_{01}+R_{02}$ defined by (\ref{R02KP2_1}) and (\ref{Kernel8R02KP2}) for real $a_1$ evidently satisfies the restriction (\ref{RealityWeakFieldKP2}) from reality condition $\overline{u}=u$ and  the restriction (\ref{BoundaryConditionWeakField}) from boundary condition (\ref{BoundaryCondition}).
This restriction satisfies more general kernel with $N$ pairs of terms of the type (\ref{R02KP2_1}) and (\ref{Kernel8R02KP2}) with real amplitudes $a_k$:
\begin{equation}\label{Kernel(KP2)4}
\fl R_0(\mu,\overline\mu;\lambda,\overline\lambda):=
R_{01}+R_{02}=\sum\limits_{k=1}^N\left(a_k\delta(\mu-\mu_{k})
\delta(\lambda-\overline{\mu}_{k})+
\overline{a}_k\delta(\mu+\overline{\mu}_{k})\delta(\lambda+\mu_{k})\right).
\end{equation}
Using the set $A_k$ of amplitudes and the sets $M_k$, $\Lambda_k$ of spectral points
\begin{eqnarray}\label{setOfAML1KP2}
\fl A_k:\,(a_1,a_1;a_2,a_2;\ldots;a_N,a_N),\quad  (k=1,\ldots,2N); \nonumber \\
\fl M_k:\,(\mu_{1},-\overline{\mu}_1;\mu_{2},-\overline{\mu}_2;\ldots;\mu_{N},
-\overline{\mu}_N),\nonumber\\
\fl \Lambda_k:\,(-\overline{\mu}_1,-\mu_{1};-\overline{\mu}_2,
-\mu_{2};\ldots;-\overline{\mu}_N,-\mu_{N}),
\end{eqnarray}
we rewrote the kernel (\ref{Kernel(KP2)4}) in the form (\ref{kernel1}). Then general formulas (\ref{MatrixN1}), (\ref{ReconstructFinal1}) give in considered case (\ref{Kernel(KP2)4}) via reconstruction formula corresponding exact
multi-soliton solution of KP-2 (\ref{ReconstructFinal1}) in general determinant form.

In the simplest case of one pair of terms in (\ref{kernel2}), i.e. for the simple kernel $R_0$ (\ref{Kernel(KP2)4}) with $N=1$, matrix $N_{kl}$ (\ref{MatrixN1}) has the form:
\begin{equation}\label{MatrixN(KP2)4}
N=\left(
  \begin{array}{cc}
    \frac{\pi}{2a_1}e^{X-iY}+\frac{1}{2\mu_{1I}}e^{-X+iY} & \frac{i}{2\mu_1}e^{\frac{F(\mu_1)-F(-\mu_1)}{2}} \\
     -\frac{i}{2\overline\mu_1}e^{\frac{F(-\overline\mu_1)-
     F(\overline\mu_1)}{2}}   &  \frac{\pi}{2a_1}e^{X+iY}+\frac{1}{2\mu_{1I}}e^{-X-iY} \\
  \end{array}
\right),
\end{equation}
here
\begin{eqnarray}\label{DeltaF(KP2)4}
\fl F(\mu_1)-F(-\mu_1) +F(-\overline\mu_1)-F(\overline\mu_1)=-2X,\nonumber\\
\fl X(x,t)=2\mu_{1I}x-8t\left(\mu^3_{1I}-3\mu_{1I}\mu^2_{1R})\right),\quad
Y(y)=4\mu_{1R}\mu_{1I}y.
\end{eqnarray}
For determinant $\det N$
\begin{equation}
\det{N}=\frac{\pi^2}{4a_1^2}e^{X}+
\frac{\mu_{1R}^2}{4\mu_{1I}^2|\mu_1|^2}e^{-X}+
\frac{\pi}{4a_1\mu_{1I}}\left(e^{-iY}+e^{iY}\right),
\end{equation}
under the requirement
\begin{equation}
\frac{\pi}{a_1}=\pm\frac{\mu_{1R}}{\mu_{1I}|\mu_1|}
\end{equation}
we derived  the following simple expression:
\begin{equation}\label{ModalphaKP_11}
\det{N}=\frac{\mu_{1R}^2}{2\mu_{1I}^2|\mu_1|^2}\left(\cosh X(x,t)+\alpha\cos Y(y)\right),\quad \alpha=\pm\frac{|\mu_1|}{\mu_{1R}},
\end{equation}
\begin{figure}[h]\label{PeriodicOnYKP2}
\begin{center}
\includegraphics[width=0.50\textwidth, keepaspectratio]{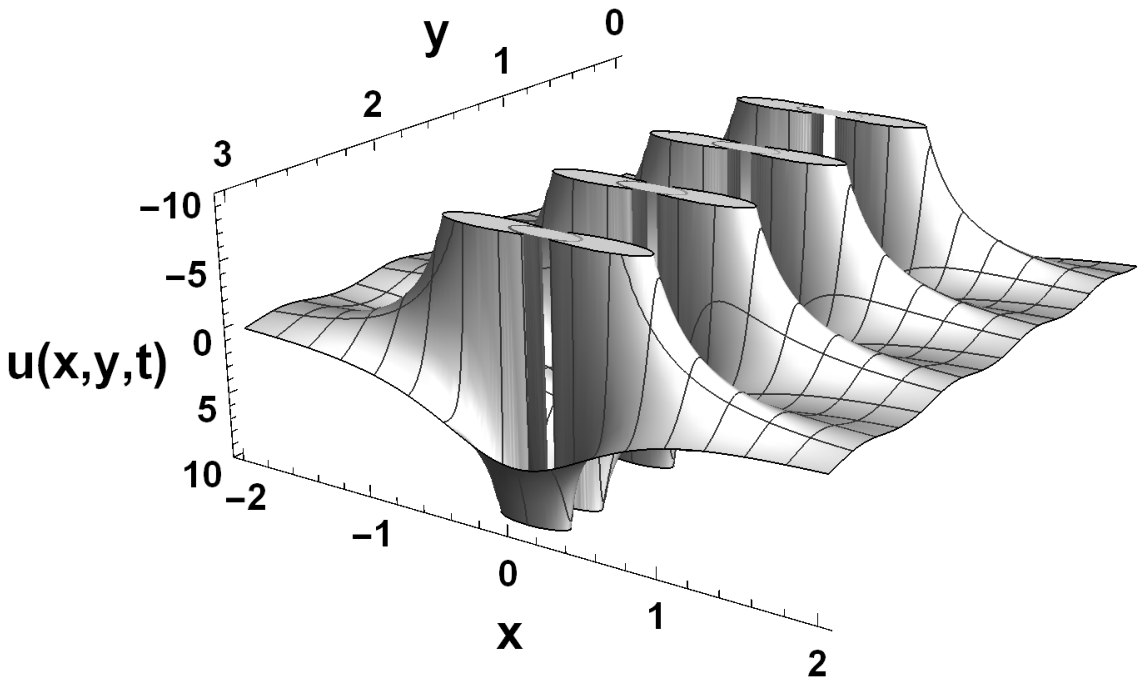}
\parbox[t]{1\textwidth}{\caption{Two-soliton solution of KP-2 $u$ (\ref{TwoSolitonKP2(1)4})  with parameter $\mu_{1I}=1$, $\mu_{1R}=2$.}\label{PeriodicOnYKP2}}
\end{center}
\end{figure}
and corresponding exact two-soliton solution that has the form:
\begin{equation}\label{TwoSolitonKP2(1)4}
u(x,y,t)=2\frac{\partial^2}{\partial x^2}\ln\det N
=8\mu^2_{1I}\frac{1+\alpha \cosh X\cos Y}{(\cosh X+\alpha \cos Y)^2}.
\end{equation}
This is exact real and due to $|\alpha|>1$ (\ref{ModalphaKP_11}) singular, periodical in $y$ solution of KP-2 equation (\ref{KP}) with integrable boundary condition (\ref{BoundaryCondition}). Boundary condition (\ref{BoundaryCondition}) is satisfied due to the fact that $u(x,y,t)$ (\ref{TwoSolitonKP2(1)4}) is even function of $y$ (through the $\cos 2Y(y)$). The considered solution propagates along $x$-axis with velocity $V_x=4(\mu^2_{1I}-3\mu^2_{1R})$.
The graph of solution (\ref{TwoSolitonKP2(1)4}) on semi-plane $y\geq 0$ is shown on figure (\ref{PeriodicOnYKP2}).

By the use of identity (\ref{Identity}) this solution can be represented as nonlinear superposition
of two complex-valued
\begin{equation}\label{Superposition1_KP2}
\fl u(x,y,t)=u_1\big|_{(\mu_1,\:\overline\mu_1)Y\rightarrow Y_D}+u_2\big|_{(-\overline\mu_1,-\mu_1)Y\rightarrow Y_D}=
\frac{2\mu_{1I}^2}{\cosh^2\frac{X+iY_D}{2}}+
\frac{2\mu_{1I}^2}{\cosh^2\frac{X-iY_D}{2}}
\end{equation}
one-solitons (\ref{u1OneSolitonKP2_11}) and (\ref{u2OneSolitonKP2_11}) with modified or deformed phase $Y_D$ defined for $\alpha>0$ by the relation:
\begin{equation}\label{DeformedPhaseY1KP2}
\cosh 2Y_{D}(y):=\alpha\cosh 2Y(y).
\end{equation}
Two terms in (\ref{Superposition1_KP2}) did not coincide with $u_1$ (\ref{u1OneSolitonKP2_11}) and  $u_2$ (\ref{u2OneSolitonKP2_11}), these terms are not the exact solutions of KP-2, but their sum is exact real two-soliton solution of KP-2 equation (\ref{KP}). The solution  (\ref{Superposition1_KP2}) corresponds to some kind of "bound state" of two one-solitons (\ref{u1OneSolitonKP2_11}) and (\ref{u2OneSolitonKP2_11}), such bound state arises as result of imposition of boundary condition $u_y\big|_{y=0}=0$ and represents \,"eigen-mode"\, of oscillations
of the field $u(x,y,t)$   in semi-plane $y\geq0$.

\section{Multi-soliton solutions of KP-2 equation with integrable boundary, pure solitonic case}
\label{Section_7}
\setcounter{equation}{0}
To pure solitonic solution of KP-2 (\ref{KP}) equation correspond the kernel $R_0(\mu,\overline\mu;\lambda,\overline\lambda)$ of $\overline\partial$-equation (\ref{di_problem1}) with pure imaginary spectral points $\mu_k=i\mu_{k0}$,  $\lambda_k=i\lambda_{k0}$, $\overline\mu_{k0}=\mu_{k0}$, $\overline\lambda_{k0}=\lambda_{k0}$. Indeed the phase $F(i\mu_{0})$ for $\overline\mu_{0}=\mu_{0}$ (\ref{F(lambda)}) is real and all exponents in $\overline\partial$-equation (\ref{di_problem1}) will be nonoscillatory. We have  the following expressions for useful phases:
\begin{eqnarray}\label{DefKP_22}
\fl F(i\mu_{k0})=-\mu_{k0}x-\mu_{k0}^2y+4t\mu_{k0}^3=\overline{F(i\mu_{k0})},\nonumber \\
\fl F(i\mu_{k0})-F(i\lambda_{k0})=-X_k(x,t)-Y_k(y),\quad F(-i\lambda_{k0})-F(-i\mu_{k0})=-X_k(x,t)+Y_k(y), \nonumber\\
\fl F(i\mu_{k0})-F(-i\mu_{k0})=-2\mu_{k0}x+8t\mu_{k0}^3,\quad F(-i\lambda_{k0})-F(i\lambda_{k0})=2\lambda_{k0}x-8t\lambda_{k0}^3,\nonumber\\
\fl F(i\mu_{k0})-F(-i\mu_{k0})+F(-i\lambda_{k0})-F(i\lambda_{k0})=-2X_k(x,t),\nonumber\\
\fl X_k(x,t):=(\mu_{k0}-\lambda_{k0})x-4t(\mu_{k0}^3-\lambda_{k0}^3) \quad Y_k(y):=(\mu_{k0}^2-\lambda_{k0}^2)y.
\end{eqnarray}

The simplest kernel
\begin{equation}\label{KernelPureSolitonKP2}
R_{01}(\mu,\overline\mu;\lambda,\overline\lambda)=
a_1\delta(\mu-i\mu_{10})\delta(\lambda-i\lambda_{10}),\quad \overline a_1=a_1
\end{equation}
leads to one-soliton solution. Matrix $N_{kl}$ for (\ref{KernelPureSolitonKP2}) due to (\ref{MatrixN1}) has the form:
\begin{equation}\label{N11_PureSoliton_KP2}
\fl N_{11}=\det{N}=\frac{\pi}{2a_1}e^{\frac{X+Y}{2}}+\frac{1}{\mu_{10}-
\lambda_{10}}e^{-\frac{X+Y}{2}}
=\frac{2}{\mu_{10}-\lambda_{10}}\cosh\frac{X+Y}{2},
\end{equation}
here for convenience the relation between amplitude $a_1$ and spectral points $\mu_{10}$, $\lambda_{10}$
\begin{equation}
  \frac{\pi}{2a_1}=\frac{1}{\mu_{10}-\lambda_{10}}
\end{equation}
is chosen. Real one-soliton solution corresponding to (\ref{N11_PureSoliton_KP2}) due to (\ref{ReconstructFinal1})  has the form:
\begin{equation}\label{u1OneSolitonKP2_2}
u_1(x,y,t)\big|_{(i\mu_{10},i\lambda_{10})}=
\frac{(\mu_{10}-\lambda_{10})^2}{2}\frac{1}{\cosh^2\left(\frac{X+Y}{2}\right)}.
\end{equation}
To the kernel $R_{02}(\mu,\overline\mu;\lambda,\overline\lambda)$
\begin{equation}\label{Kernel22PureSolitonKP2}
R_{02}(\mu,\overline\mu;\lambda,\overline\lambda):=
a_1\delta(\mu+i\lambda_{10})\delta(\lambda+i\mu_{10}),
\end{equation}
corresponds another real one-soliton solution, which can be obtained from (\ref{u1OneSolitonKP2_2}) by the change $\mu_{10}\leftrightarrow - \lambda_{10}$:
\begin{equation}\label{u2OneSolitonKP2_2}
u_2(x,y,t)\big|_{(-i\lambda_{10},-i\mu_{10})}=
\frac{(\mu_{10}-\lambda_{10})^2}{2}
\frac{1}{\cosh^2\left(\frac{X-Y}{2}\right)}.
\end{equation}
The solutions $u_1$ (\ref{u1OneSolitonKP2_2}) and  $u_2$ (\ref{u2OneSolitonKP2_2}) are useful for constructions of present section.

The sum $R_{01}+R_{02}$ of  kernels  (\ref{KernelPureSolitonKP2}) and (\ref{Kernel22PureSolitonKP2}) or more general
 kernel $R_0$ with $N$ pairs of terms, similar to (\ref{KernelPureSolitonKP2}) and (\ref{Kernel22PureSolitonKP2})
\begin{equation}\label{Kernelsum(KP2)PureSoliton}
\fl R_0(\mu,\overline\mu;\lambda,\overline\lambda)=\sum\limits^N_{k=1}
(a_k\delta(\mu-i\mu_{k0})\delta(\lambda-i\lambda_{k0})+
a_k\delta(\mu+i\lambda_{k0})\delta(\lambda+i\mu_{k0})),\quad \overline{a}_k=a_k,
\end{equation}
satisfies restrictions (\ref{RealityWeakFieldKP2}) and (\ref{BoundaryConditionWeakField}) from reality $\overline u=u$ and boundary (\ref{BoundaryCondition}) conditions and leads to $N$ pairs of bounded with each other solitons of the type (\ref{u1OneSolitonKP2_2})
and (\ref{u2OneSolitonKP2_2}).
Using the set $A_k$ of amplitudes and the sets $M_k$, $\Lambda_k$ of spectral points
\begin{eqnarray}\label{setOfAML1KP22}
\fl A_k:\,(a_1,a_1;a_2,a_2;\ldots;a_N,a_N), (k=1,\ldots,2N); \nonumber \\
\fl M_k:\,(i\mu_{10},-i\lambda_{10};i\mu_{20},-i\lambda_{20};
\ldots;i\mu_{N0},-i\lambda_{N0}),\nonumber\\
\fl\Lambda_k:\,(i\lambda_{10},-i\mu_{10};i\lambda_{20},-i\mu_{20};
\ldots;i\lambda_{N0},-i\mu_{N0}),
\end{eqnarray}
we rewrote the kernel (\ref{Kernelsum(KP2)PureSoliton}) in the form (\ref{kernel1}); then  general determinant formula (\ref{ReconstructFinal1}) gives exact real $2N$-soliton solution $u(x,y,t)$ of KP-2 equation.

In the simplest case of $N=1$, one pair of terms in $R_0$ (\ref{Kernelsum(KP2)PureSoliton}),
general formulas (\ref{MatrixN1}) and (\ref{ReconstructFinal1}) give corresponding to $R_0$ (\ref{Kernelsum(KP2)PureSoliton}) real exact two-soliton solution of KP-2 equation (\ref{KP}). Matrix $N_{kl}$ (\ref{MatrixN1}) for $R_0$ (\ref{Kernelsum(KP2)PureSoliton}) has the form:
\begin{equation}\label{MatrixN(KP2)5}
N=\left(
  \begin{array}{cc}
    \frac{\pi}{2a_1}e^{\frac{X+Y}{2}}+\frac{1}{\mu_{10}-
    \lambda_{10}}e^{-\frac{X+Y}{2}} & \frac{1}{2\mu_{10}}e^{{\frac{F(i\mu_{10})-F(-i\mu_{10})}{2}} }\\
     \frac{1}{2\lambda_{10}}e^{\frac{F(-i\lambda_{10})-
     F(i\lambda_{10})}{2}}     &  \frac{\pi}{2a_1}e^{\frac{X-Y}{2}}+\frac{1}{\mu_{10}-\lambda_{10}}
     e^{-\frac{X-Y}{2}} \\
  \end{array}
\right).
\end{equation}
For $\det N$ we derived
\begin{equation}
\det{N}=\frac{\pi^2}{4a_1^2}e^{X}+\frac{(\mu_{10}+\lambda_{10})^2}
{4\lambda_{10}^2\mu_{10}^2(\mu_{10}-\lambda_{10})^2}e^{-X}
+\frac{\pi}{2a_1(\mu_{10}-\lambda_{10})}\left(e^{Y}+e^{-Y}\right).
\end{equation}
Imposing on amplitude $a_1$ and $\lambda_{10}$, $\mu_{10}$ for convenience the relation
\begin{equation}
\frac{\pi^2}{a_1^2}=\frac{(\mu_{10}+
\lambda_{10})^2}{\lambda_{10}\mu_{10}(\mu_{10}-\lambda_{10})^2}, \quad \lambda_{10}\mu_{10}>0,
\end{equation}
we simplified the expression for $\det N$:
\begin{equation}
\det{N}=
\frac{(\mu_{10}+\lambda_{10})^2}{2\mu_{10}\lambda_{10}
(\mu_{10}-\lambda_{10})^2}\left(\cosh X+\alpha\cosh Y\right),
\end{equation}
here due to definitions (\ref{DefKP_22})
\begin{eqnarray}
X(x,t)=(\mu_{10}-\lambda_{10})x-4t(\mu_{10}^3-\lambda_{10}^3), \quad
Y(y)=(\mu_{10}^2-\lambda_{10}^2)y,\nonumber\\
\alpha:=\pm\frac{2\sqrt{\mu_{10}\lambda_{10}}}{(\mu_{10}+\lambda_{10})},
\quad |\alpha|\leq1.
\end{eqnarray}
\begin{figure}[h]\label{PureSolitonicKP2}
\begin{center}
\includegraphics[width=0.50\textwidth, keepaspectratio]{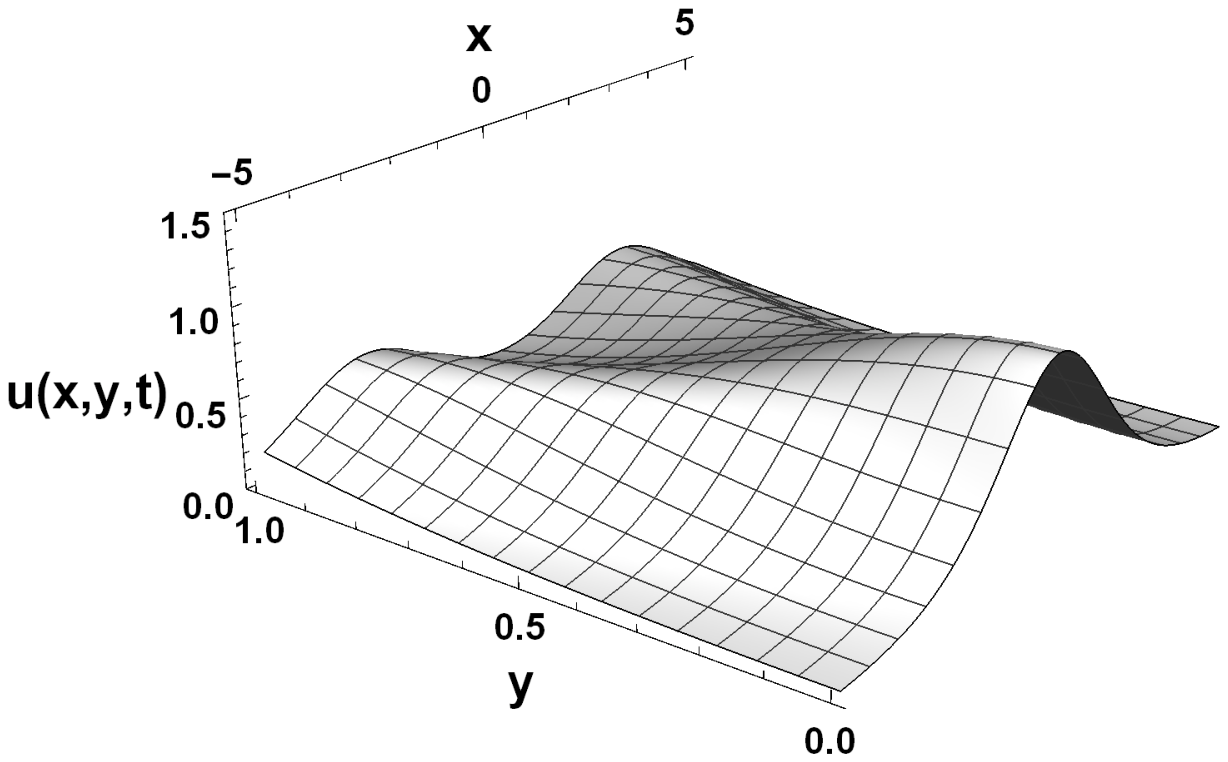}
\parbox[t]{1\textwidth}{\caption{Two-soliton solution of KP-1 $u$ (\ref{TwoSolitonKP2(1)5})  with parameter $\lambda_{10}=1$, $\mu_{10}=2$.}\label{PureSolitonicKP2}}
\end{center}
\end{figure}
Corresponding exact real two-soliton solution has the form:
\begin{equation}\label{TwoSolitonKP2(1)5}
u(x,y,t)=2\frac{\partial^2}{\partial x^2}\ln\det N=2(\mu_{10}-\lambda_{10})^2\frac{1+\alpha \cosh X\cosh Y}{(\cosh X+\alpha \cosh Y)^2}.
\end{equation}
The solution (\ref{TwoSolitonKP2(1)5}) is real and evidently for $|\alpha|<1$ nonsingular two-soliton solution of KP-2 equation with integrable boundary condition (\ref{BoundaryCondition}). Boundary condition (\ref{BoundaryCondition}) is satisfied due to the fact that $u(x,y,t)$ (\ref{TwoSolitonKP2(1)5}) is even function of $y$ (through the $\cosh Y(y)$). The considered solution propagates along $x$-axis with velocity $V_x=4(\mu^2_{10}+\mu_{10}\lambda_{10}+\lambda^2_{10})$.
The graph of solution (\ref{TwoSolitonKP2(1)5}) on semi-plane $y\geq 0$ is shown on figure (\ref{PureSolitonicKP2}).

Using the identity (\ref{Identity}) and introducing for $\alpha>0$ instead of $Y(y)$ modified phase $Y_D(y)$
\begin{equation}\label{DeformedPhaseY2KP2}
\cosh Y_{D}(y):=\alpha\cosh Y(y),
\end{equation}
we obtained the exact two-soliton solution (\ref{TwoSolitonKP2(1)5})
as nonlinear superposition of one-soliton solutions $u_1(x,y,t)$ (\ref{u1OneSolitonKP2_2}) and  $u_2(x,y,t)$ (\ref{u2OneSolitonKP2_2})
\begin{eqnarray}\label{Superposition2_KP2}
u(x,y,t)=u_1\big|_{(i\mu_{10},\:i\lambda_{10}) Y\rightarrow Y_D}+u_2\big|_{(-i\lambda_{10},\:-i\mu_{10}) Y\rightarrow Y_D}=\nonumber\\
=\frac{(\mu_{10}-\lambda_{10})^2}
{2\cosh^2\left(\frac{X+Y_D}{2}\right)}+
\frac{(\mu_{10}-\lambda_{10})^2}{2\cosh^2\left(\frac{X-Y_D}{2}\right)}
\end{eqnarray}
 with modified (or deformed) phase $Y\rightarrow Y_D$. Two terms in (\ref{Superposition2_KP2}) did not coincide with $u_1$ (\ref{u1OneSolitonKP2_2}) and  $u_2$ (\ref{u2OneSolitonKP2_2}), these terms are not the exact solutions of KP-2, but their sum is exact real two-soliton solution of KP-2 equation (\ref{KP}). In some sense  (\ref{Superposition2_KP2}) corresponds to bound state of two one-solitons (\ref{u1OneSolitonKP2_2}) and (\ref{u2OneSolitonKP2_2}), such bound state arises as result of imposition of boundary condition $u_y\big|_{y=0}=0$ and represents \,"eigen-mode"\, state
of the field $u(x,y,t)$   in semi-plane $y\geq0$.

\section{Multi-soliton solutions of KP-2 equation with integrable boundaries, periodical on $X(x,t)$ variables}
\label{Section_8}
\setcounter{equation}{0}
Next, we construct multi-soliton solution of KP-2 equation (\ref{KP})  corresponding to the kernel $R_0(\mu,\overline\mu;\lambda,\overline\lambda)$ of $\overline\partial$-equation (\ref{di_problem1}) with real spectral points $\overline\mu_k=\mu_{k}$,  $\overline\lambda_k=\lambda_{k}$. In this case the phase $F(\mu_{k})$ (\ref{F(lambda)}) is real on $y$ variable and pure imaginary on $x$, $t$ variables:
\begin{equation}
F(\mu_{k})=i(\mu_{k}x+4t\mu_{k0}^3)+\mu_{k}^2y=
\overline{F(-\mu_{k})}.
\end{equation}
For phases $F(\mu_{k})-F(\lambda_l)$ in the matrices $A_{kl}$ (\ref{A&tildeA}) and $N_{kl}$ (\ref{MatrixN1}) we derived the following useful expressions:
\begin{eqnarray}\label{XYKP-21}
\fl F(\mu_k)-F(\lambda_k)=i\left((\mu_k-\lambda_k)x+4t(\mu_k^3-\lambda_k^3)
\right)+(\mu_k^2-\lambda_k^2)y:=iX_k+Y_k,\nonumber \\
\fl  X_k=(\mu_k-\lambda_k)x+4t(\mu_k^3-\lambda_k^3),\quad Y_k(y)=(\mu_k^2-\lambda_k^2)y, \nonumber\\
\fl  F(-\lambda_k)-F(-\mu_k)=iX_k-Y_k;\quad
 F(\mu_k)-F(-\mu_k)+F(-\lambda_k)-F(\lambda_k)=2iX_{k}.
\end{eqnarray}
The simplest kernel
\begin{equation}\label{KernelPeriodicalXKP2}
R_{01}(\mu,\overline\mu;\lambda,\overline\lambda)=
a_1\delta(\mu-\mu_{1})\delta(\lambda-\lambda_{1}),\quad \overline\mu_1=\mu_1\ ,
\overline\lambda_1=\lambda_1
\end{equation}
leads evidently to complex-valued  one-soliton solution. Matrix $A_{kl}$, due to (\ref{A&tildeA}) and (\ref{XYKP-21})
for the choice $\frac{2a_1}{\pi(\mu_1-\lambda_1)}=1$, has the form:
\begin{equation}\label{AKP-2_1}
\fl  A_{11}:=\det{A} =1 +\frac{2ia_1}{\pi(\mu_1-\lambda_1)}e^{F(\mu_1)-F(\lambda_1)} =2e^{\frac{i (X_1+\frac{\pi}{2})+Y_1}{2}}\cos\left(\frac{X_1+\frac{\pi}{2}-iY_1}{2}\right),
\end{equation}
and one-soliton solution calculated by the use of reconstruction formula (\ref{ReconstructFinal1}) and (\ref{AKP-2_1}) has the form:
\begin{equation}\label{u1OneSolitonKP2_3}
u_1(x,y,t)\big|_{(\mu_1,\lambda_1)}=2\frac{\partial^2}{\partial x^2}\ln\det A=
-\frac{(\mu_1-\lambda_1)^2}{2}
\frac{1}{\cos^2\left(\frac{X_1+\frac{\pi}{2}-iY_1}{2}\right)}.
\end{equation}
To the kernel $R_{02}(\mu,\overline\mu;\lambda,\overline\lambda)$
defined by the formula
\begin{equation}\label{Kernel2PureSolitonKP2}
R_{02}(\mu,\overline\mu;\lambda,\overline\lambda):=R_{01}(-\lambda,-\overline\lambda;-\mu,-\overline\mu)=
a_1\delta(\mu+\lambda_1)\delta(\lambda+\mu_1),
\end{equation}
corresponds to complex one-soliton solution $u_2(x,y,t)$ which can be obtained from $u_1$ by the change $\mu_1\leftrightarrow-\lambda_1$:
\begin{equation}\label{u2OneSolitonKP2_3}
u_2(x,y,t)\big|_{(-\lambda_1,-\mu_1)}=-\frac{(\mu_1-\lambda_1)^2}{2}
\frac{1}{\cos^2\left(\frac{X_1+\frac{\pi}{2}+iY_1}{2}\right)}=\overline{u_1(x,y,t)\big|_{\mu_1,\lambda_1}}.
\end{equation}
These simple one-soliton solutions (\ref{u1OneSolitonKP2_3}) and (\ref{u2OneSolitonKP2_3}) are used throughout the section.
The sum $R_{01}+R_{02}$ of kernels (\ref{KernelPeriodicalXKP2}) and (\ref{Kernel2PureSolitonKP2}), or more general kernel  with $N$ pairs of terms of considered type
\begin{equation}\label{Kernel(KP2)PeriodicalX}
R_0(\mu,\overline\mu;\lambda,\overline\lambda):=\sum\limits_{k=1}^N\left(a_k\delta(\mu-\mu_{k})\delta(\lambda-\lambda_{k})+
a_k\delta(\mu+\lambda_k)\delta(\lambda+\mu_k)\right)
\end{equation}
satisfies the restriction (\ref{BoundaryConditionWeakField}) from boundary condition (\ref{BoundaryCondition}). The kernel (\ref{Kernel(KP2)PeriodicalX}) using  the sets $A_k$ of amplitudes and $M_k$, $\Lambda_k$ of spectral points
\begin{eqnarray}
A_k:\,(a_1,a_1;a_2,a_2;\ldots;a_N,a_N), (k=1,...,2N); \nonumber \\
M_k:\quad (\mu_1,-\lambda_1;\mu_2,-\lambda_2,\ldots,\mu_N,-\lambda_N), \nonumber\\
\Lambda_k:\quad (\lambda_1,-\mu_1;\lambda_2,-\mu_2,\ldots,\lambda_N,-\mu_N)
\end{eqnarray}
can be rewritten in general form (\ref{kernel1}):
\begin{equation}\label{Kernel(KP2)PeriodicalX1}
R_0(\mu,\overline\mu;\lambda,\overline\lambda):=
\sum\limits_{k=1}^{2N}A_k\delta(\mu-M_{k})\delta(\lambda-\Lambda_{k}),
\end{equation}
and via general formulas (\ref{ReconstructFinal}) and (\ref{MatrixN1}) leads
to exact 2N-soliton solutions of KP-2 equation.

As an example we constructed the simplest two-soliton solution of this type. Matrix (\ref{MatrixN})
\begin{equation}
N_{kl}:=\frac{\pi }{2a_l}e^{\frac{-F(\mu_k)+F(\lambda_l)}{2}}A_{kl}
\end{equation}
 for $N=1$ case of kernel $R_0$ (\ref{Kernel(KP2)PeriodicalX}) with one pair of terms of the type  (\ref{KernelPeriodicalXKP2})  and (\ref{Kernel2PureSolitonKP2})
has the following form:
\begin{equation}\label{MatrixN(KP2)6}
N:=
i\left(
  \begin{array}{cc}
    \frac{\pi}{2ia_1}e^{-\frac{iX_{1}+Y_1}{2}}
    +
    \frac{1}{\mu_{1}-\lambda_{1}}
    e^{\frac{iX_{1}+Y_1}{2}} & \frac{1}{2\mu_1}e^{\frac{F(\mu_1)-F(-\mu_1)}{2}}
     \\
    -\frac{1}{2\lambda_1}
    e^{\frac{F(-\lambda_1)-F(\lambda_1)}{2}}  &  \frac{\pi}{2ia_1}e^{-\frac{iX_{1}-Y_1}{2}}
    +\frac{1}{\mu_{1}-
    \lambda_{1}}e^{\frac{iX_{1}-Y_1}{2}}  \\
  \end{array}
\right).
\end{equation}
For $\det N$ (\ref{MatrixN(KP2)6}) we obtained:
\begin{equation}\label{detN(KP2)}
\fl \det{N}=\left(\frac{\pi^2}{(2a_1)^2}e^{-i(X_1+\frac{\pi}{2})}-
\frac{(\mu_1+\lambda_1)^2}{4\lambda_1\mu_1(\mu_1-\lambda_1)^2}
e^{i(X_1+\frac{\pi}{2})}+\frac{i\pi}{2a_1(\mu_1-\lambda_1)}
\left(e^{-Y_1}+e^{Y_1}\right)\right).
\end{equation}
Required in (\ref{detN(KP2)})
\begin{equation}
\lambda_1\mu_1>0,\quad
\frac{\pi^2}{(a_1)^2}=\frac{(\mu_1+\lambda_1)^2}{\lambda_1\mu_1(\mu_1-\lambda_1)^2}
\Rightarrow\frac{\pi}{a_1}=\pm\frac{(\mu_1+\lambda_1)}{\sqrt{\lambda_1\mu_1}(\mu_1-\lambda_1)},
\end{equation}
we derived simple expression for  for $\det N$:
\begin{equation}
\det{N}=i\frac{(\mu_{1}+\lambda_1)^2}
{2(\mu_1-\lambda_1)^2\mu_{1}\lambda_1} \left(\cos{(X_1+\frac{\pi}{2})}+\alpha\cosh{Y_1}\right)
\end{equation}
with
\begin{equation}
\fl X_1=(\mu_1-\lambda_1)x+4t(\mu_1^3-\lambda_1^3),\quad Y_1(y)=(\mu_1^2-\lambda_1^2)y, \quad \alpha=\pm\frac{2\sqrt{\mu_1\lambda_1}}{(\mu_1+\lambda_1)},\quad |\alpha|\leq1.
\end{equation}
\begin{figure}[h]\label{PeriodicOnXTKP2}
\begin{center}
\includegraphics[width=0.50\textwidth, keepaspectratio]{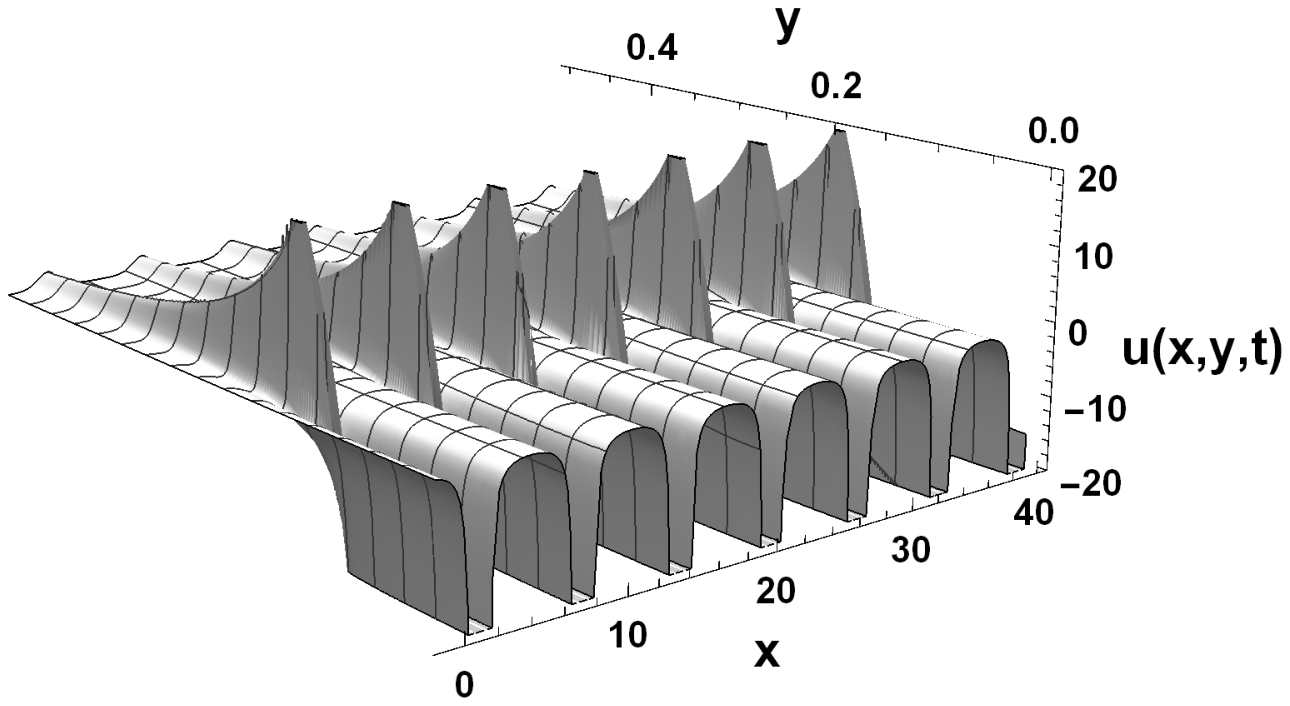}
\parbox[t]{1\textwidth}{\caption{Two-soliton solution of KP-2 $u$ (\ref{TwoSolitonKP2(1)6})  with parameter $\lambda_{1}=1$, $\mu_{1}=2$.}\label{PeriodicOnXTKP2}}
\end{center}
\end{figure}
The reconstruction formula (\ref{ReconstructFinal}) gives finally the exact real two-soliton solution:
\begin{equation}\label{TwoSolitonKP2(1)6}
u(x,y,t)=2\frac{\partial^2}{\partial x^2}\ln\det N=-2(\mu_1-\lambda_1)^2{\frac{1+\alpha \cos(X_1+\frac{\pi}{2})\cosh Y_1}
{(\cos(X_1+\frac{\pi}{2}) + \alpha \cosh Y_1)^2}}.
\end{equation}
The solution (\ref{TwoSolitonKP2(1)6}) is real and, due to $|\alpha|\leq1$, singular two-soliton solution of KP-2 equation with integrable boundary condition (\ref{BoundaryCondition}). Boundary condition (\ref{BoundaryCondition}) is satisfied due to the fact that $u(x,y,t)$ (\ref{TwoSolitonKP2(1)6}) is even function of $y$ (through the $\cosh Y_1(y)$). The considered solution, periodic in the phase $X_1(x,t)$, propagates along $x$-axis with velocity $V_x=-4(\mu^2_{1}+\mu_1\lambda_1+\lambda^2_{1})$.
The graph of solution (\ref{TwoSolitonKP2(1)6}) on semi-plane $y\geq 0$  is shown on figure (\ref{PeriodicOnXTKP2}).

Using the identity (\ref{Identity}) we rewrote real two-soliton solution (\ref{TwoSolitonKP2(1)6}) in the form of the sum of two  simple one-soliton solutions $u_1$ (\ref{u1OneSolitonKP2_3}) and $u_2$ (\ref{u2OneSolitonKP2_3})
\begin{eqnarray}\label{Superposition3_KP2}
u(x,y,t)=u_1\big|_{(\mu_1,\:\lambda_1)\: Y\rightarrow Y_D}+u_2\big|_{(-\lambda_1,\:-\mu_1)\: Y\rightarrow Y_D}=\nonumber\\
=-\frac{(\mu_1-\lambda_1)^2}{2\cos^2\left(\frac{X_1+\frac{\pi}{2}-iY_D}{2}\right)}-
\frac{(\mu_1-\lambda_1)^2}{2\cos^2\left(\frac{X_1+\frac{\pi}{2}+iY_D}{2}\right)}
\end{eqnarray}
with modified or deformed phase $Y(y)\rightarrow Y_D(y)$ defined for $\alpha>0$ by the relation:
\begin{equation}\label{DeformedPhaseY3KP2}
\alpha\cosh Y(y):=\cosh Y_{D}(y).
\end{equation}
It should be emphasised that two complex-valued terms $u_1\big|_{(\mu_1,\:\lambda_1)\: Y\rightarrow Y_D}$ and $u_2\big|_{(-\lambda_1,\:-\mu_1)\: Y\rightarrow Y_D}$  in (\ref{Superposition3_KP2}), given via (\ref{u1OneSolitonKP2_3}) and (\ref{u2OneSolitonKP2_3})  by the change $Y\rightarrow Y_D$ and
taken separately, are not exact solutions of KP-2 equation; but their sum  is real exact solution of KP-2.

We can also state that  exact two-soliton solution (\ref{TwoSolitonKP2(1)6}) corresponds to \,"bound state"\, of two simple complex-valued one-solitons  (\ref{u1OneSolitonKP2_3}) and (\ref{u2OneSolitonKP2_3}). The effect of interaction of these one-solitons (\ref{u1OneSolitonKP2_3}) and (\ref{u2OneSolitonKP2_3}) consists in the nonlinear change of the phase $Y\rightarrow Y_D$.
Two-soliton solution (\ref{TwoSolitonKP2(1)6}) (or (\ref{Superposition3_KP2}))  resembles certain eigenmode of oscillations of field $u(x,y,t)$ in semi-plane $y\geq 0$ arising as result of imposition of boundary condition (\ref{BoundaryCondition}) on this field.

\section{Conclusions}
\label{Section_9}
\setcounter{equation}{0}

In the present paper we derived new classes of exact real multi-soliton solutions of KP-1,2 equations with integrable boundary condition $u_{y}\big|_{y=0}=0$
and developed general scheme for this in framework of $\overline\partial$-dressing method. We demonstrated that reality $u=\overline u$ and boundary conditions for solutions $u$ can be effectively satisfied exactly and this
leads to some restrictions on parameters of solutions, i.e. on amplitudes $A_k$ and spectral points $\mu_k$, $\lambda_k$ of delta-form kernel $R_0$ of $\overline\partial$-problem (\ref{di_problem1}).

We presented the simplest examples  of real two-soliton exact solutions (nonsingular and singular) as illustrations. These calculated in the paper solutions  belong to the class of solutions with integrable boundary conditions.The imposition on the field $u$ of boundary condition (\ref{BoundaryCondition}) leads to formation of bounded with each other simple one-solitons, the eigenmodes of the field $u(x,y,t)$  on semi-plane $y\geq 0$. Such eigenmodes of coherently connected with each other simple solitons  propagate with some velocity along $x$-axis.

We also demonstrated the effectiveness of $\overline\partial$-dressing in calculations of multi-soliton solutions with integrable boundary condition.
The developed in present paper procedure for calculation via $\overline\partial$-dressing of new classes of exact real multi-soliton solutions can be effectively applied to all other integrable (2+1)-dimensional nonlinear equations, these research is now in progress and  corresponding results will be published elsewhere.

An exciting physical applications of calculated in the present paper exact multi-soliton solutions of KP equation should be noted.  KP equation can be applied for description of fluids flows in thin films on inclined surfaces in Earth gravity field. There may be, due to some specific experimental boundary conditions, some kind of fluid \,"excitations"\, in such films, periodical or solitonic waves, that could be observed by hydrodynamics experimentalists.

\section*{References}
\setcounter{equation}{0}


\begin{thebibliography}{99}

\bibitem{KadPetv} B. B. Kadomtsev, V. I. Petviashvili, On the stability of solitary waves in weakly dispersing media, Dokl. Akad. Nauk SSSR, 192:4 (1970), 753-756.

\bibitem{Dryuma} V. S. Dryuma, Analitic solution of the two-dimemsional Korteveg-de Vries (KdV) equation, Sov. Phys. JETF Lett., 19 (1974), 387-388.

\bibitem{ZakharovShabat} V.E.Zakharov, A.B.Shabat. A scheme for integrating the nonlinear equations of mathematical physics by the method of the inverse scattering problem I.  Funct Anal Its Appl (1974) 8: 226. https://doi.org/10.1007/BF01075696

\bibitem{ZakharovShabat2} V.E.Zakharov, A.B.Shabat. Integration of nonlinear equations of mathematical physics by the method of inverse scattering II.  Funct Anal Its Appl (1979) 13: 166. https://doi.org/10.1007/BF01077483

\bibitem{Manakov} S.V. Manakov. The inverse scattering transform for the time-dependent Schrodinger equation and Kadomtsev-Petviashvili equation // Physica D. 1981. Vol. 3 (1-2),  pp. 420-427. doi: 10.1016/0167-2789(81)90145-7

\bibitem{AblowitzSegur} M.J.Ablowitz, H.Segur. Solitons and the Inverse Scattering Transform. Series: SIAM Studies in Applied Mathematics. Society for Industrial and Applied Mathematics 1981.

\bibitem{Fokas&Ablowitz}    A.S. Fokas, M.J. Ablowitz, The inverse scattering transform for multidimensional (2+1) problems // Lecture Notes in Physics, vol. 189, p.137-183 Nonlinear Phenomena. [Proceedings of the CIFMO School and Workshop held at Oaxtepec, Mexico November 29 - December 17, 1982.]

\bibitem{NovikovManakov} S.P. Novikov, S.V. Manakov, L.P. Pitaevskii, V.E. Zakharov. Theory of Solitons: The Inverse Scattering Method. Series: Monographs in Contemporary Mathematics. Springer US 1984.

\bibitem{AblowitzClarkson} M.J.Ablowitz, P.A. Clarkson. Solitons, Nonlinear Evolution Equations and Inverse Scattering. London Mathematical Society Lecture Note Series. Cambridge University Press, 1991.

\bibitem{KonopelchenkoBook1} B.G. Konopelchenko. Introduction to Multidimensional Integrable Equations: The Inverse Spectral Transform in 2+1 Dimensions. New York: Plenum Press, 1992.

\bibitem{KonopelchenkoBook2} B.G. Konopelchenko. Solitons in Multidimensions: Inverse Spectral Transform Method. Singapore: World Scientific, 1993.

\bibitem{Zakharov&Manakov} V. E. Zakharov, S. V. Manakov, Construction of higher-dimensional nonlinear integrable systems and of their solutions, Funktsional. Anal. i Prilozhen., 19:2 (1985), 11-25; Funct. Anal. Appl., 19:2 (1985), 89-101. https://doi.org/10.1007/BF01078388

\bibitem{Zakharov} V.E. Zakharov.  Commutating operators and nonlocal  problem // Plasma theory and Nonlinear and turbulent processes in Physics / Ed. by Erokhin N.S., Zakharov V.E., Sitenko A.G., Chernousenko V.M. Bar'yakhtar V.G. Kiev: Naukova Dumka, 1988. Vol. 1, pp. 152.

\bibitem{Bogdanov&Manakov}  L.V. Bogdanov, S.V.Manakov. The non-local  problem and (2+1)-dimensional soliton equations // Journal of Physics A. 1988. Vol. 21 (10), pp. 537-544. doi:10.1088/0305-4470/21/10/001


\bibitem{Zakharov90} V.E. Zakharov On the dressing method // Inverse Methods in Action / Ed. By P.C.Sabatier. Springer, 1990, pp. 602.


\bibitem{Beals&Coifman1} R. Beals, R.R. Coifman. The D-bar approach to inverse scattering and nonlinear evolutions// Physica D. 1986. Vol. 18 (1-3), pp. 242-249. doi: 10.1016/0167-2789(86)90184-3

\bibitem{Beals&Coifman2} R. Beals, R.R. Coifman.  Linear spectral problems, non-linear equations and the $\overline\partial$-method. Inverse Problems. 1989. Vol. 5 (87), pp. 87-130. doi: 10.1088/0266-5611/5/2/002

\bibitem{Sklyanin} E. K. Sklyanin, Boundary conditions for integrable equations,  Funktsional. Anal. i Prilozhen., 21:2 (1987),  86-87; Funct. Anal. Appl., 21:2 (1987), 164-166. https://doi.org/10.1007/BF01078038

\bibitem{Habibullin}  E. V. Gudkova, I. T. Habibullin, Kadomtsev-Petviashvili Equation on the Half-Plane, Theoret. and Math. Phys., 140:2 (2004), 1086-1094, https://doi.org/10.1023/B:TAMP.0000036539.35565.1f

\bibitem{Habibullin2}   Habibullin I. T., Gudkova E. V., Boundary Conditions for Multidimensional Integrable Equations, Funktsional. Anal. i Prilozhen., 38:2 (2004), 71-83; Funct. Anal. Appl., 38:2 (2004), 138-148. https://doi.org/10.4213/faa109

\bibitem{HabibullinSinGSchr} I.T. Khabibullin, Boundary conditions for nonlinear equations compatible with integrability, Theor Math Phys (1993) 96: 845. https://doi.org/10.1007/BF01074113

\bibitem{HabibullinSineG}  I.T. Khabibullin, Sine-Gordon equation on the semi-axis. Theor Math Phys (1998) 114: 90. https://doi.org/10.1007/BF02557111

\bibitem{HabibullinAdlerShabat} V.E. Adler, L.T. Habibullin, A.B. Shabat, Boundary value problem for the KDV equation on a half-line. Theor Math Phys (1997) 110: 78.
https://doi.org/10.1007/BF02630371

\bibitem{HabibullinKDV} I.T. Khabibullin, KDV equation on a half-line with the zero boundary condition. Theor Math Phys (1999) 119: 712. https://doi.org/10.1007/BF02557381

\bibitem{Vereshchagin} V. L. Vereshchagin, Integrable boundary conditions for (2+1)-dimensional models of mathematical physics. Theor Math Phys (2012) 171: 792.  https://doi.org/10.1007/s11232-012-0075-9

\bibitem{FokasBook} A.S. Fokas. A Unified Approach to Boundary Value Problems. Society for Industrial and Applied Mathematics. 2008. doi:10.1137/1.9780898717068.



\bibitem{Dubrovsky&Topovsky&OstreinovKP} V.G. Dubrovsky, A.V. Topovsky, G.M. Ostreinov. The construction of exact solutions of Kadomtsev-Petviashvili (KP-2) equation with integrable boundary conditions via dibar-dressing method. Doklady Akademii nauk vysshei shkoly Rossiiskoi Federatsii - Proceedings of the Russian higher school Academy of sciences, 2018, no. 4 (41), pp. 7-29. doi: 10.17212/1727-2769-2018-4-7-29.



\end{thebibliography}
\end{document}